\documentclass[prb,twocolumn,superscriptaddress,showpacs]{revtex4}
\usepackage{amsmath}
\usepackage{graphicx}
\usepackage{latexsym}

\begin{document}

\renewcommand{\Im}{{\rm Im \,}}
\newcommand{\up}{\uparrow}
\newcommand{\down}{\downarrow}

\renewcommand{\vr} {{\bf r}}
\newcommand{\vR} {{\bf R}}
\newcommand{\rem}[1]{}

\title[local SIC]{Self-interaction correction in multiple scattering theory}

\author{M. L{\"u}ders}
\affiliation{Daresbury Laboratory, Daresbury, Warrington, WA4 4AD, UK}
\author{A. Ernst}
\affiliation{Max Planck Institut f{\"u}r Mikrostrukturphysik, Weinberg 2, D-06120 Halle, Germany}
\author{M. D{\"a}ne}
\affiliation{Fachbereich Physik,
  Martin-Luther-Universit\"at Halle-Wittenberg,Friedemann-Bach-Platz 6, D-06099 Halle, Germany}
\affiliation{Daresbury Laboratory, Daresbury, Warrington, WA4 4AD, UK}
\author{Z. Szotek}
\affiliation{Daresbury Laboratory, Daresbury, Warrington, WA4 4AD, UK}
\author{A. Svane}
\affiliation{ Institute of Physics and Astronomy, University of Aarhus, DK-8000 Aarhus, Denmark}
\author{D. K{\"o}dderitzsch}
\affiliation{Fachbereich Physik,
  Martin-Luther-Universit\"at Halle-Wittenberg,Friedemann-Bach-Platz 6, D-06099 Halle, Germany}
\affiliation{Daresbury Laboratory, Daresbury, Warrington, WA4 4AD, UK}
\author{W. Hergert}
\affiliation{Fachbereich Physik,
  Martin-Luther-Universit\"at Halle-Wittenberg,Friedemann-Bach-Platz 6, D-06099 Halle, Germany}
\author{B. L. Gy\"orffy}
\affiliation{H.H. Wills Physics Laboratory, University of Bristol, Tyndall
  Avenue, Bristol BS8 1TL, UK}
\author{W. M. Temmerman}
\affiliation{Daresbury Laboratory, Daresbury, Warrington, WA4 4AD, UK}
\date{\today}


\begin{abstract}
  We propose a simplified version of self-interaction corrected local
  spin-density (SIC-LSD) approximation, based on multiple scattering
  theory, which implements self-interaction correction locally, within
  the KKR method.  The multiple scattering aspect of this new SIC-LSD
  method allows for the description of crystal potentials which vary 
  from site to site in a random fashion and the calculation of
  physical quantities averaged over ensembles of such potentials
  using the coherent potential approximation (CPA).  
  This facilitates applications of the
  SIC to alloys and pseudoalloys which could describe disordered local
  moment systems, as well as intermediate valences.  As a
  demonstration of the method, we study the well-known
  $\alpha$-$\gamma$ phase transition in Ce, where we also explain how
  SIC operates in terms of multiple scattering theory.
\end{abstract}

\maketitle

\section{Introduction}

The self-interaction corrected local spin density (SIC-LSD)
approximation \cite{PZ81,HeatonEtAl:83} has proved to be a
useful scheme to describe static correlations in strongly correlated
electron systems.  In particular, it can determine whether an electron is
delocalized or localized, i.e., whether its orbital is part of
the valence states or not.  This leads to a determination of the
number of valence states and a nominal valence, as demonstrated by
numerous calculations on rare earths, actinides, transition metal
oxides, including the parent compounds of the high T$_{c}$ materials
and the CMR materials.
\cite{StrangeEtAl:99,SvaneEtAl:00,TemmermanEtAl:01,SciencePetit:03,SzotekEtAl:03,
TyerEtAl:04,BanachTemmerman:04}

The full SIC-LSD scheme is unfortunately difficult to implement.\cite{Brisbane} 
This is due to the repeated transformations from reciprocal space (k-space)
to real space to evaluate the self-interaction potential and the back
transformations to k-space to solve the band structure problem. So far
most applications of the full SIC-LSD formalism have been implemented in
the LMTO-ASA (linearized muffin-tin orbitals) band structure method.\cite{Andersen:75} 
In this paper a simpler but more versatile scheme is developed and 
implemented within multiple scattering theory, in the Korringa, Kohn,
Rostocker (KKR) formulation. 
Its main advantage, thanks to a straightforward determination of
the Green's function, is a possible generalization to alloys via the 
coherent potential approximation (CPA).\cite{Soven:67, StocksEtAl:78, GyorffyStocks:79, 
FaulknerStocks:81, StocksWinter:84}
Since a single-site approximation underpins this new formulation, in what follows 
it is referred to as a local self-interaction correction (LSIC) formalism.
It is based on the experience with
the full SIC-LSD implementation showing that to better 
than 98\% the electron is localized on the site under consideration, 
which justifies the single-site approximation.
Whilst in the full LMTO-ASA implementation, the representation of the
localized orbitals over a real-space cluster determines the extent of
these orbitals, in the present scheme the degree of localization 
is determined by the energy dependence of the single-site phase shift,
in particular the width of its resonance, corresponding to the
localized electron. A broader resonance would imply reduced
localization.

The SIC-LSD formalism for solids has been developed into a scheme
that treats both localized and delocalized electrons on equal footing.
The decision whether an electron should be considered as localized
or delocalized is based on a delicate balance between the
energy gain due to the inclusion of the self-interaction correction 
(localization) energy
and the energy loss in band or hybridization energy.\cite{StrangeEtAl:99}
While this methodology has been successful in differentiating
localized from delocalized electrons, i.e., a dual character of the
electron, it does not describe the interesting crossover between
localized and delocalized states which occurs for example in heavy
fermion systems. 
Our aim here is to develop a theory which describes local fluctuations 
of the electronic configurations between that where an electron can 
be said to be localized and another where an electron is delocalized. 
It will be shown that the present local formulation of SIC-LSD readily  
lends itself to be the basic idea of such a development.
The origin of our approach goes back to the invention and use of the 
coherent potential approximation to describe the charge and spin 
fluctuations about the Hartree-Fock solution of the Hubbard model by 
Hubbard himself (Hubbard III approximation).\cite{Hubbard1, Hubbard2, Hubbard3} 
The present implementation of this idea 
rests on its generalization to account for the corresponding fluctuations 
about the local density approximation (LDA) to the first-principles  
density functional theory (DFT).\cite{HK, KS, DFT} In the literature this 
generalization, when applied to spin fluctuations, 
is referred to as the KKR-CPA implementation of the disordered local moment 
(DLM) picture.\cite{DLM1, DLM2} Indeed, the present work can be considered 
as the further elaboration of this basic idea in which the LSIC replaces the  
LDA as the local description of the electronic structure and the attention 
is being focused on the valance fluctuations.
Interestingly, it is now well established that Hubbard's so-called "alloy 
analogy" approximation, which prompted the use of the CPA, leaves out of 
consideration some very important fluctuations. The most significant of 
these are those which give rise to a Kondo-like resonance at the Fermi 
energy in the case of the Hubbard model and correspond to such a
qualitatively new physics as the Mott transition. The relevance of this 
in the present context is that such fluctuations are well described by 
the dynamical mean field theory (DMFT)\cite{DMFT} whose static limit, for 
the Hubbard model, is precisely the "alloy analogy" CPA  of 
Hubbard.\cite{Hubbard3} This point was particularly clearly explained in 
the recent paper of Kakahashi.\cite{Kakehashi} Consequently, it is reasonable 
to regard our LSIC based KKR-CPA-DLM calculations as investigations of the 
static limit of a  yet undeveloped first principles DMFT. In what follows 
when we refer to the need to include dynamical effects in the theory it is 
the above theoretical considerations we will have in mind.

The paper is organized as follows. In Sec. \ref{sec:picture} we
outline the physical picture underlying the present approach. In Sec.
\ref{sec:formalism} a general formulation of SIC-LSD, following
Perdew and Zunger,\cite{PZ81} is briefly summarized with reference to some
aspects of the LMTO-ASA implementation based on the Wannier function representation
of localized orbitals.\cite{Brisbane} In Sec. \ref{sec:formalism2}, 
the formalism of the local self-interaction corrected local spin density
(LSIC-LSD) within multiple scattering theory is described in detail. 
There we concentrate on the phase shifts and single-site Green's function
from which the SIC charge and potential, corresponding to localized
electron states, are calculated within the KKR method.
Since the latter can be easily extended to include coherent potential
approximation, section \ref{sec:CPA} briefly summarizes its most
important equations in terms of the multiple scattering
quantities. In section \ref{sec:FT} the formalism is extended to
finite temperatures. The potential and versatility of the LSIC method is
demonstrated on the application to the $\alpha-\gamma$ phase transition 
in Ce. In section \ref{sec:Ce} we first discuss the f-phase shifts, 
total energies, lattice parameters, densities of states (DOS), and
spectral functions at T=0K for the $\alpha-$ and $\gamma-$ phases, as 
obtained from the LSIC-KKR method. Wherever appropriate we compare with 
the results of the full SIC-LSD
implementations within LMTO-ASA.\cite{Brisbane} In this section we also 
present calculations for finite temperatures and the full phase diagram of
the $\alpha-\gamma$ phase transition. Both the CPA and DLM are
utilized to accomplish the latter, and to illustrate how the
present approach is capable to describe both spin and valence fluctuations
at finite temperatures. Section \ref{sec:discussion} is devoted to various
aspects of the present approach and among them a 
consideration of how intermediate valence could be realized within
the present implementation, which motivates a possible
generalization to include dynamics, as outlined in section \ref{sec:outlook}. 
The paper is summarized in section \ref{sec:conclusions}.

\section{Physical Picture}
\label{sec:picture}

In the present formulation of SIC, we adopt the physical picture of multiple-scattering
theory, where a solid is represented by an array of non-overlapping scattering centers.
The electronic motion is then broken down into a sequence of scattering events and a
free propagation in between. 
The most useful concept of this method is a phase shift, describing 
scattering of electrons from ions, the scattering centers in a solid. 
If a phase shift is resonant it is reminiscent of a bound state at positive
energies, i.e., above the zero of the potential which in our case is
the muffin-tin zero.  The energy derivative of the phase shift is
related to the Wigner delay-time.  If this is large the electron will
spend a long time on the site.  Such 'slow' electrons will be much
more affected by the spurious self-interaction and therefore should
see a SI-corrected potential.

In most systems, where the electrons are truly delocalized, the
self-interaction contribution to the potential is negligible and
therefore the LDA is an excellent approximation. 
When the phase shift has a resonance one has to calculate the
self-interaction correction for this $(l,m)$ angular momentum channel.
This is accomplished by calculating the one-electron charge
density for this channel, defining the charge density for the
self-interaction correction. From this one can readily calculate the
self-interaction potential which has to be added to the LSD potential,
and then the new phase shifts are calculated for the total (SIC-LSD) 
potential.
This has to be carried out $m$ channel by $m$ channel for a given angular
momentum $l$.  The minimum of the total energy will determine the
optimum configuration of ($l$,$m$) channels to be self-interaction
corrected.  Therefore to each of the $m$ channels one can assign
two potential functions, V$^{\rm SIC-LSD}_{\rm eff}$(r) and V$_{\rm
  eff}^{\rm LSD}$(r). 
The formalism determining the energy functional associated with the
potential V$^{\rm SIC-LSD}_{\rm eff}$(r) is briefly outlined in the 
next section. 
It should be mentioned here that if the total energies corresponding 
to these two different 
potentials are sufficiently close, one can envisage dynamical effects 
to play an important role as a consequence of possible tunneling between 
these states. We shall return to this point in the later sections.


\section{SIC-LSD Formalism}
\label{sec:formalism}

It has been pointed out by Perdew and Zunger\cite{PZ81} that density
functional theory (DFT) schemes, like the local spin density
approximation, suffer from a spurious self-interaction of the electrons
with themselves. In principle, this self-interaction term should
vanish exactly, as it does in the Hartree-Fock theory. In practice,
however, this cancellation is incomplete.  Perdew and Zunger suggested
an approximate solution to this problem, which was constructed for
finite systems but is here extended to solids in a different way as compared
to previous implementations for solids.\cite{HeatonEtAl:83}

The usual representation of the total energy within the LSD-DFT
formalism in the Kohn-Sham approach\cite{KS} is
\begin{eqnarray}
  \label{eq:Etot_LDA}
  E^{\rm LSD}[n_{\uparrow},n_{\downarrow}] &=& \sum_{\alpha\sigma}^{\rm occ} \langle
  \phi_{\alpha\sigma}|-\nabla^2|\phi_{\alpha\sigma} \rangle + E_{\rm ext} \nonumber \\
 && +  E_{\rm H}[n]+E_{\rm xc}^{\rm LSD}[n_{\uparrow},n_{\downarrow}] ,
\end{eqnarray}
where  $\phi_{\alpha\sigma}$'s are the Kohn-Sham orbitals, ($\alpha\sigma$ is a
combined index labeling the orbital and spin ($\uparrow$ or $\downarrow$), 
respectively), $n_{\alpha\sigma}=|\phi_{\alpha\sigma}|^2$,
$n_{\sigma}= \sum^{\rm occ}_{\alpha} n_{\alpha\sigma}$,
$n=n_{\uparrow}+n_{\downarrow}$. $E_{\rm ext}$ is the external field energy functional,
$E_{\rm H}$ is the Hartree energy
\begin{equation}
  \label{eq:E_H}
  E_{\rm H}[n]= \int \! {\rm d}^3 \! r \! \int \! {\rm d}^3 r'  \,
  \frac{n (\mathbf{r}) n (\mathbf{r}')}{|\mathbf{r}-\mathbf{r}'|} \, ,
\end{equation}
and $E_{\rm xc}^{\rm LSD}$ is the LSD approximation to the exchange-correlation
energy functional.
On the basis of the above, Perdew and Zunger proposed a
self-interaction corrected LSD on an orbital by orbital basis
\begin{eqnarray}
  \label{eq:PZ}
  E^{\rm SIC-LSD}[\{n_{\alpha\sigma}\}]
  &=& E^{\rm LSD}[n_{\uparrow},n_{\downarrow}] - \nonumber \\
  &-& \sum_{\alpha\sigma}^{\rm occ} (E_{\rm H}[n_{\alpha\sigma}] +
E_{\rm xc}^{\rm LSD}[n_{\alpha\sigma},0]),
\end{eqnarray}
by subtracting explicitly the self-Coulomb and self-exchange and self-correlation energy of
all \emph{occupied} orbitals. This correction restores the property
that the true functional $E[n]$ should have, namely that the
self-Coulomb energy exactly cancels the self-exchange and self-correlation energy for every single
orbital, $E_{\rm H}[n_{\alpha\sigma}]+E^{\rm exact}_{\rm
  xc}[n_{\alpha\sigma},0]=0$.  This leads to an orbital dependent
SIC-potential seen by an electron in orbital $\phi_{\alpha\sigma}$,
\begin{eqnarray}
  \label{eq:SIC-pot}
  V^{\rm SIC-LSD}_{{\rm eff},\alpha\sigma}  (\mathbf{r}) &=& 
  \underbrace{V_{\rm ext}(\mathbf{r}) + V_{\rm H}[n](\mathbf{r}) + 
  V_{\rm xc \sigma}^{\rm LSD}[n_{\up},n_{\down}] (\mathbf{r})}_{V^{\rm LSD}_{\rm eff, \sigma}(\mathbf{r})} \nonumber \\
  && - \underbrace{V_{\rm H}[n_{\alpha \sigma}](\mathbf{r})
     - V^{\rm LSD}_{{\rm xc},\sigma}[n_{\alpha \sigma},0] (\mathbf{r})
}_{V^{\rm SIC}(\mathbf{r})} ,
\end{eqnarray}
with the external lattice potential $V_{\rm ext}({\bf r})$, and
\begin{eqnarray}
  \label{eq:vHartree}
  V_{\rm H}[n](\mathbf{r}) &=& 
  2 \int \! {\rm d}^3 \! r' \,  
  \frac{n(\mathbf{r}')}{|\mathbf{r}-\mathbf{r}'|} \, , \\
  V^{\rm LSD}_{\rm{xc}, \sigma}[n_{\up},n_{\down}](\mathbf{r}) &=& 
    \frac{\delta E^{\rm LSD}_{\rm xc}[n_{\up},n_{\down}]}{\delta n_{\sigma}} .
\end{eqnarray}

This self-interaction correction vanishes exactly only for extended
states.  In order to apply the SIC scheme to solids, the approach by
Perdew and Zunger has to be generalized.  This involves simultaneously
a Wannier representation of the orbitals, necessary to determine
$n_{\alpha \sigma}$ of Eq.~(\ref{eq:SIC-pot}), and a Bloch representation to
solve the band structure problem. Furthermore, the Wannier functions
are required to fulfill the localization criterion which ensures that
the energy functional is stationary with respect to unitarian mixing
among the orbitals. This localization criterion is necessary, because
the SIC is not invariant under unitary transformations of the occupied
orbitals.  This is in contrast with the LSD where a unitary
transformation of the occupied orbitals leaves the LSD potential
invariant, since the total charge density remains unaltered.  For the
orbital dependent SIC potential V$^{\rm SIC}$ such a unitary
transformation will change V$^{\rm SIC}$. The localization criterion
$<\alpha|V^{\rm SIC}_{\alpha}-V^{\rm SIC}_{\beta}|\beta>=0$ determines the
unitary transformation which ensures the global minimum of the total
energy and the hermiticity of the Hamiltonian.  Solutions of this
equation usually take the form of the eigenvector $|\alpha>$ having
weight in one channel only $(({\rm Im} \alpha_{j})^{2}+({\rm Re}
 \alpha_{j})^{2}=1)$ which would be different from the channel where
the weight of the eigenvector $|\beta>$ $(({\rm
  Im} \beta_{i})^{2}+({\rm Re} \beta_{i})^{2}=1)$ is concentrated,
i.e., $i$ is not equal to $j$.  This generalization forms the basis of
the SIC implementations\cite{Brisbane,St-Odile} which
start from a band-picture scenario.

\section{Single-site SIC-LSD formalism}
\label{sec:formalism2}

As already mentioned, the proposed generalization of the Perdew and Zunger idea 
is based on the notion of resonances in scattering theory, which are
the reminiscence of atomic states in the solid. Core states are
represented by bound states at negative energies, where the
imaginary part of the generalized complex phase shift jumps abruptly by $\pi$. Localized
valence states still have very sharp resonances but band-like states
are characterized by slowly varying phase shifts.

The central quantity of (scalar-relativistic)
multiple scattering theory is the single-particle Green's function\cite{FaulknerStocks:81}
\begin{eqnarray}
\label{GF-KKR}
G_\sigma(\vr,\vr';\epsilon) &=& 
\sum_{L L'} \bar{Z}^i_{L\sigma}(\vr_i;\epsilon) \, \tau^{i j}_{\sigma LL'}(\epsilon) \,
Z^j_{L'\sigma}(\vr_j';\epsilon) \nonumber \\
&& - \sum_L \bar{Z}^i_{L\sigma}(\vr_<;\epsilon) J^i_{L\sigma}(\vr_>;\epsilon) \delta_{i j},
\end{eqnarray}
with $\vr = \vR_i +\vr_i$, where $\mathbf{r}_i$ is a vector inside the
cell at $\mathbf{R}_i$, $L=(l,m)$ denotes the combined index for the
decomposition into symmetrized lattice harmonics $Y_{L}$ and
$\mathbf{r}_<$($\mathbf{r}_>$) is the vector smaller (larger) in
magnitude from the pair ($\mathbf{r},\mathbf{r}'$). The building
blocks of the Green's function are the regular and irregular solutions
of the radial Schr{\"o}dinger equation at a given (complex) energy
$\epsilon$
\begin{eqnarray}
Z^i_{L\sigma}(\vr_i;\epsilon) &=& Z^i_{l\sigma}(r_i;\epsilon) Y_L(\hat{\mathbf{r}}_i) \\
\bar{Z}^i_{L\sigma}(\vr_i;\epsilon) &=& Z^i_{l\sigma}(r_i;\epsilon) Y^*_L(\hat{\mathbf{r}}_i) \\
J^i_{L\sigma}(\vr_i;\epsilon) &=& J^i_{l\sigma}(r_i;\epsilon) Y_L(\hat{\mathbf{r}}_i) .
\end{eqnarray}
The scattering-path matrix $\underline{\tau}$ ( in $L,L'$ and $\sigma$ representation)
\begin{equation}
\label{tau-mat}
\underline{\tau}(\epsilon) = 
\left[ \underline{t}^{-1}(\epsilon) - \underline{g}(\epsilon) \right]^{-1}
\end{equation}
is related to the structural Green's function $\underline{g}(\epsilon)$,
describing the free propagation between the scattering centers, and the
$\underline{t}$ matrix defines the single site scattering.

The total valence charge density per spin $\sigma$ is given by:
\begin{equation}
\label{eq:density-G}
n_\sigma(\vr) = -\frac{1}{\pi}
\int_{E_{B}}^{E_F} \! \! {\rm d} \epsilon \,
\Im G_\sigma(\vr,\vr;\epsilon) ,
\end{equation}
where $E_{B}$ and $E_{F}$ denote the bottom of the valence band and
the Fermi energy, respectively. In standard LSD calculations, the new
effective potential for the next iteration of the self-consistency
cycle is calculated from this density (now including the core contributions) as
\begin{equation}
  \label{eq:V_eff_lsd}
  V^{\rm{LSD}}_{\rm{eff},\sigma}(\vr) = V_{\rm
    ext}(\vr) + V_{\rm H}[n](\vr) + V^{\rm LSD}_{\rm{xc}, \sigma}[n_{\uparrow},n_{\downarrow}](\vr).
\end{equation}
In order to remove the spurious self-interaction, still present in
this potential, we consider the problem of electrons moving in an
array of scatterers.  As already mentioned, an electron which shows
localized behavior has a sharp resonance in its phase shift, associated
with a large Wigner-delay time on a particular site.
%
%
To determine the SIC charge we will consider for a moment the atomic
limit, i.e., the situation where the scatterers are far apart.  In this
case the single site $t$-matrix and the local multiple scattering
$\tau$-matrix coincide, and all occupied states correspond to bound
states.
In this limit each bound state contributes exactly the charge of one
electron, and this charge can be calculated by integrating the
diagonal of the spectral function just around the energy of the bound
state.  In order to be able to decompose the charge density
(Eq.~(\ref{eq:density-G})) into different angular momentum channels,
we choose symmetry adapted spherical harmonics. These are defined by
applying a unitary transformation to the ordinary real (or complex)
spherical harmonics, such that the on-site scattering matrix becomes
diagonal,
\begin{equation}
\sum_{L_1, L_2} \, U^{\dagger}_{L L_1} \tau^{i i}_{L_1,L_2}(\epsilon) \, U_{L_2 L'} 
= \delta_{L L'} \tilde{\tau}^{i i}_{L L}(\epsilon) \, .
\end{equation}
It is easy to verify that the required transformation matrix $U$ is, in fact,
independent of the energy $\epsilon$.
This transformation to symmetry adapted spherical harmonics also
ensures that the degeneracy of states, which are localized, is 
conserved. We will demonstrate this later by SI-correcting the
triplet states (one by one) of the Ce $f$-manifold.
In this symmetrized representation, the Green's function, which in 
the atomic limit equals the atomic Green's function, becomes
diagonal with respect to this quantum number.  Hence we can decompose
the spin resolved charge density into its $L$ components and define
the charge of a state, characterized by its principle quantum number $n$,
angular momentum $L$ and spin $\sigma$:
\begin{equation}
n^{\rm SIC}_{n L \sigma}(\vr) = -\frac{1}{\pi}
\int_{E_1}^{E_2} \! \! {\rm d} \epsilon \,
\Im G_{L L, \sigma} (\vr,\vr;\epsilon) ,
\end{equation}
where the energies $E_1$ and $E_2$ lie slightly below
and above the energy of the state $n L \sigma$.
Within the multiple scattering formulation, in the atomic limit, this
charge density can be written as
\begin{eqnarray}
  \label{eq:nL-SIC}
n^{{\rm SIC}}_{iL\sigma}(\vr) &=& 
-\frac{1}{\pi} \int_{E_1}^{E_2} \! \! {\rm d} \epsilon \, 
\Im \Bigg[ \bar{Z}^i_{L\sigma}(\vr_i;\epsilon) \tau^{i i}_{\sigma L L}(\epsilon)
Z^i_{L\sigma}(\vr_i;\epsilon) \nonumber \\
&&- \bar{Z}^i_{L\sigma}(\vr_i;\epsilon) J^i_{L\sigma}(\vr_i;\epsilon) \Bigg] ,
\end{eqnarray}
where $i$ is the site index, since in this case the single site
$t$-matrix and the $\tau$-matrix are obviously identical. This, of
course, is not the case for a solid with finite lattice spacings.
When considering resonances in a solid it is not a priori clear
whether to use the $t$- or $\tau$-matrix for calculating the
charge density in question. The main difference between using the $t$-
matrix or $\tau$-matrix is that the latter does include a small
hybridization of the localized state with the surrounding atoms, while
the former does not.
Also the choice of the lower and upper integration limits is not
clearly defined. We will now give a short discussion of the possible
modes for calculating the SIC charge of a resonant state. 

The lower integration limit is most reasonably chosen to be the bottom
of the energy contour $E_{\rm B}$. However, care has to be taken that this
contour always encloses the SI-corrected states. Nota bene, in the case of Ce,
discussed in the following sections, the contour also includes the $5p$
semi-core states.  The upper integration limit could be either chosen
such that the SIC charge density integrates to exactly one electron,
or simply set to the Fermi energy $E_{\rm F}$.  Using the $t$-matrix,
we find that we have to integrate up to an $E_{\rm top}$, which is
slightly above the Fermi level in order to capture one electron. The
$\tau$-matrix, on the other hand, due to hybridization, yields a
charge of one for energies slightly below the Fermi energy.
Unfortunately, when dealing with the $\tau$-matrix, it is computationally
very expensive to assume $E_2$ different from the Fermi energy. However,
the excess charge due to integrating up to the Fermi level is only of the
order of a hundredth of an electron. (The missing charge in case of
integrating the $t$-matrix up to the Fermi level is of similar magnitude.)
In the following we used the $\tau$-matrix integrated up to $E_{\rm
  F}$ to determine the SIC charge. Some tests with the $t$-matrix, and
the requirement of a SIC-charge of unity, resulted in an upward shift
of the total energies by about 1 mRy. 

The charge density, calculated in either of the proposed ways, is used
to construct the effective self-interaction free potential, namely
\begin{equation}
  \label{eq:V_L_SIC}
V_{{\rm eff},iL\sigma}^{{\rm SIC-LSD}}(\vr) =
V_{\rm eff,\sigma}^{\rm LSD}(\vr)
 - V_{\rm H}[n_{iL\sigma}^{\rm SIC}](\vr) - V_{\rm xc}^{\rm LSD}[n_{iL\sigma}^{\rm SIC},0](\vr) .
\end{equation}
In this paper we only consider the spherically symmetric part of the
SIC density and SIC potential. Hence, the $t$-matrix is diagonal
in $l$ and $m$.  Here it should be noted that, if we transform the
equations back to the unsymmetrized (real or complex) spherical 
harmonics, this effective potential assumes matrix character with
respect to the angular momentum, and would not simply couple to the density,
but rather to the non-diagonal $lm,l'm'$ density matrix. This is
conceptually analogous to the rotationally invariant formulation
of LDA+U by Dudarev {\em et al.}\cite{DLC97}

For each self-interaction corrected channel
$\tilde{L} = (\tilde{l},\tilde{m})$ and $\tilde{\sigma}$, we replace the $\tilde{L}-{th}$
element of the original $t$-matrix by the one obtained from the
SI-corrected potential
\begin{equation}
  \label{eq:t_L_SIC}
\tilde{t}^i_{L\sigma} = t^i_{L\sigma} (1 - \delta_{L,\tilde{L}} \delta_{\sigma,\tilde{\sigma}}) 
+ t^{i,{\rm SIC-LSD}}_{\tilde{L}\sigma} \delta_{L,\tilde{L}} \delta_{\sigma,\tilde{\sigma}},
\end{equation}
where $t^i_{L\sigma}$ is the $t$-matrix calculated from the effective
potential $V^{\rm LSD}_{{\rm eff},\sigma}(\vr)$, and $t^{i,{\rm
 SIC-LSD}}_{L\sigma}$ is calculated from the SI-corrected potential
$V_{{\rm eff},iL\sigma}^{{\rm SIC-LSD}}(\vr)$. This $\tilde{t}$-matrix is
then used in Eq.~(\ref{tau-mat}) to calculate the new, SI-corrected,
scattering path matrix $\tilde{\underline{\tau}}$. From the latter the
new SIC-LSD charge density is calculated, and the process is iterated
until self-consistency is reached.
The correction term, which approximately compensates the
self-repulsion, is an attractive potential which will pull down in
energy the state to which it is applied (see section \ref{sec:Ce}).

To finish this section we would like to mention that in contrast to
the LSD, the SIC-LSD Hamiltonian is not invariant under unitary
transformations of the occupied orbitals.  As pointed out before, in
the full implementation the localization criterion is applied to make
the solution stationary under this unitary mixing of states.  In the
present implementation there is no such localization criterion, and
one has to be solely guided by the energetics to find the global energy
minimum.  Note that the total energies are invariant under a
rotation of the coordinate system owing to the symmetry adapted spherical harmonics 
that diagonalize the $\tau$-matrix at the $\Gamma$ point.  Hence,
the energies of the configurations where, one by one, each state out
of a degenerate manifold is localized are the same. This was tested on
Ce by SI-correcting all $f$-states separately. As expected, the
energies for localizing any of the $T_{1u}$ (or respectively the
$T_{2u}$) states were identical.

\section{CPA generalization}
\label{sec:CPA}

One of the advantages of the multiple scattering implementation of the
SIC-LSD formalism is that it can be easily generalized to include the
coherent potential approximation, \cite{Soven:67, StocksEtAl:78, GyorffyStocks:79,
FaulknerStocks:81, StocksWinter:84}
extending the range of applications to
random alloys. In addition, one can use it to treat static correlations
beyond LSD by studying pseudoalloys whose constituents are composed e.g.
of two different states of a given system: one delocalized, described by
the LSD potential, and another localized, corresponding to the SIC-LSD
potential.
Combined with the DLM formalism for spin-fluctuations,\cite{DLM1,DLM2} this 
allows also for different orientations of the local moments of the
constituents involved.

In the CPA extension of the SIC-LSD formalism, bearing in mind its
single-site aspect, it is required to satisfy the following CPA
self-consistency condition

\begin{equation}
\label{eq:tau_CPA}
c \underline{\tau}^{A,00}(\epsilon) + (1 -c) \underline{\tau}^{B,00}(\epsilon)  
=  \underline{\tau}^{C,00}(\epsilon), \\
\end{equation}

where the impurity $\tau-$matrices $\underline{\tau}^{A,00}(\epsilon)$ and
$\underline{\tau}^{B,00}(\epsilon)$ are given by

\begin{eqnarray}
\label{eq:tau_AB}
\underline{\tau}^{A,00}(\epsilon) 
&=& \frac{\underline{\tau}^{C,00}(\epsilon)} 
{(1 + \underline{\tau}^{C,00}(\epsilon) (\underline{t}_{A}(\epsilon) - \underline{t}_{C}(\epsilon)))} \\
\underline{\tau}^{B,00}(\epsilon) 
&=& \frac{\underline{\tau}^{C,00}(\epsilon)} 
{(1 + \underline{\tau}^{C,00}(\epsilon) (\underline{t}_{B}(\epsilon) - \underline{t}_{C}(\epsilon)))}
\, \, ,
\end{eqnarray}

\noindent
and the $\tau-$matrix of the coherent potential approximation

\begin{equation}
\label{eq:tau_C}
\underline{\tau}^{C,00}(\epsilon)  =  \frac{1} {\Omega_{BZ}} \int d^{3}k \frac{1} 
{(\underline{t}^{-1}_{C}(\epsilon) - \underline{g}(\vec{k},\epsilon))} .
\end{equation}

\noindent
Here $\Omega_{BZ}$ is the volume of the Brillouin zone (BZ),
$\underline{t}_{A}(\epsilon)$ and $\underline{t}_{B}(\epsilon)$ are
the respective single site scattering matrices of the $A$ and $B$
species, occurring with the concentrations $c$ and $1-c$, respectively, 
and $\underline{t}_{C}(\epsilon)$ is the $t-$matrix of the
effective CPA medium. Note that in the CPA extension of the
SIC-LSD formalism, the CPA condition (Eq.~(\ref{eq:tau_CPA})) is an additional 
self-consistency criterion to the usual charge or potential self-consistency.

\noindent
Finally, it should be mentioned that the formalism of this section can be 
easily generalized from a binary to
a multi-component case, as described in Ref.~\onlinecite{Tem-Pin}. In addition,
an extention of the LSIC-CPA formalism to finite temperatures can be implemented
as described in the next section (\ref{sec:FT}).

\section{Finite temperatures} 
\label{sec:FT}

\noindent
In this section we summarize the relevant formulas underlying the 
finite temperature generalization of the present formalism in
its CPA extention.
In contrast to $T=0$, at finite temperatures the
physics is dominated by thermal (classical) fluctuations. Therefore,
to properly take into account the finite temperature effects, one needs
to evaluate the free energy of the system (alloy) under consideration, namely

\begin{eqnarray}
F(T,c,V) &=& E_{\rm tot}(T,c,V)  - T \Big( S_{\rm el}(T,c,V) \nonumber \\
&& + S_{\rm mix}(c)  + S_{\rm mag}(c) + S_{\rm vib}(c) \Big) ,
\end{eqnarray}

\noindent
where $S_{\rm el}$ is the electronic (particle-hole) entropy, $S_{\rm mix}$
the mixing entropy, $S_{\rm mag}$ the magnetic
entropy, and $S_{\rm vib}$ the entropy originating from the lattice
vibrations.

\noindent
The electron-hole entropy is defined as\cite{NicholsonEtAl:94}
\begin{eqnarray}
\label{eq:1}
S_{\rm el}(T,c,V) &=& - k_B \int \! {\rm d} \epsilon \, n(\epsilon)
\Big( f_{\beta}(\epsilon) \ln f_{\beta}(\epsilon) \nonumber \\
&& + (1-f_{\beta}(\epsilon)) \ln (1-f_{\beta}(\epsilon)) \Big) ,
\end{eqnarray}
where $k_B$ is the usual Boltzmann constant and $f_{\beta}(\epsilon)$
denotes the Fermi-Dirac distribution function. The entropy of mixing
in the case of a binary system can be expressed as
\begin{equation}
\label{eq:4}
S_{\rm mix}(c) = - k_B (c \ln c + (1-c) \ln (1-c) ).
\end{equation}
The magnetic and vibrational entropies are strongly dependent on the 
system under consideration, and they will be discussed in more detail 
in the section dedicated to the phase diagram of Ce. 

Finally, note that in the definition of the free energy, the finite temperature 
enters only via the Fermi-Dirac distribution and the entropy contributions,
while for the exchange-correlation energy, being part of the total
energy E$_{\rm tot}$, the T=0K LDA (LSD)
approximation is used for all temperatures, which is a common
practice in all {\it ab initio} calculations.

This section completes the formal description and implementation of the
LSIC-KKR-CPA band structure method. In the following sections we shall
illustrate the potential and versatility of this approach for describing
strongly correlated electron systems by an application to Ce. There
we present both the T=0K and finite temperature results, including the 
phase diagram of the famous $\alpha - \gamma$ phase transition.

\section{Ce $\alpha - \gamma$ phase transition} 
\label{sec:Ce}

%

Ce is the first element in the periodic table that contains an 
$f$ electron, and shows an
interesting phase diagram.\cite{KoskenmakiGschneidner:78} In
particular, its isostructural (fcc $\to$ fcc) $\alpha-\gamma$ phase
transition is associated with a 15\%-17\% volume collapse and
quenching of the magnetic moment.\cite{KoskenmakiGschneidner:78} The
low-pressure $\gamma$-phase shows a local magnetic moment, and is
associated with a trivalent configuration of Ce. At the temperatures
in which the $\gamma$-phase is accessible, it is in a paramagnetic
disordered local moment state.  Increasing the pressure, the material
first transforms into the $\alpha$-phase, which is indicated to be in an
intermediate valence state with quenched magnetic moment. At high
pressures (50 kbar at room temperature) Ce eventually transforms into
the tetravalent $\alpha'$-phase.  With increasing temperature, the
$\alpha-\gamma$ phase transition shifts to higher pressures, ending in
a critical point (600K, 20 kbar), above which there is a continuous
crossover between the two phases.

In the following paragraphs we first discuss the SIC and non-SIC $f-$ phase
shifts and densities of states. Then we compare the results of LSIC total
energies, for the ferromagnetic arrangement of local moments, with the
earlier calculations of the full SIC implementation, in order to
benchmark the new method. After discussing the density of states of
the LDA and SIC-LSD calculations, we mix the two phases using
the CPA and DLM for the spins. Finally, allowing for finite temperatures,
we describe the full phase diagram of Ce.

\subsection{f-Phase shifts and corresponding densities of states}
\label{subsec:PhaseShifts}

Before presenting our results for the phase diagram, we discuss briefly
the scattering properties of a single Ce site.
In particular, we concentrate on the phase shifts and corresponding
densities of states for $f$ electrons.
\begin{figure}
\includegraphics[scale=0.32]{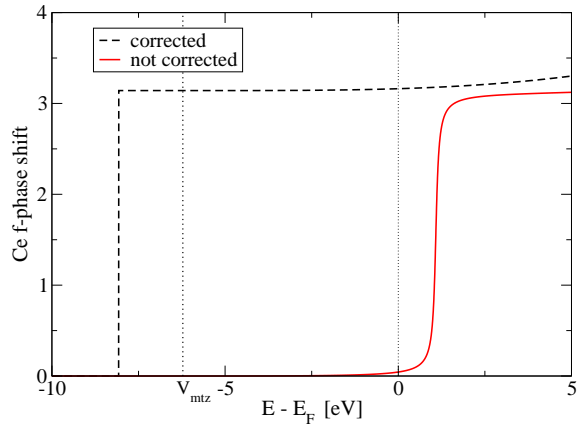}
\caption{
\label{fig:phase-shifts} (color online) 
Phase shifts of the SI-corrected and uncorrected $f$-states in Ce from the SIC-LSD calculation. 
The energies are relative to the Fermi level.}
\end{figure}
\begin{figure}
\includegraphics[scale=0.32]{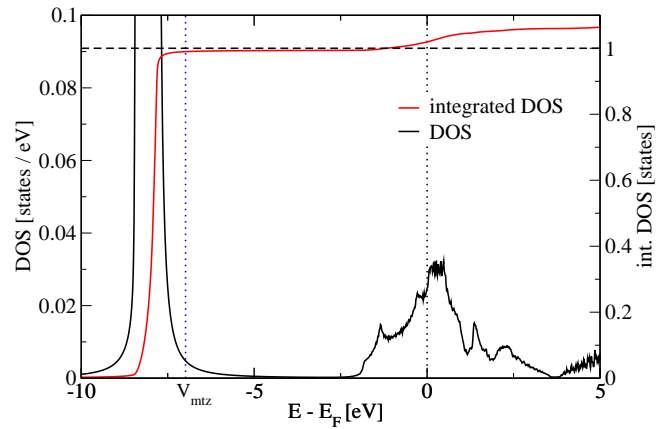}
\caption{
\label{SIC-charge} (color online)
Density of states and integrated density of states of the SI-corrected $f$-channel.}
\end{figure}
In Fig. \ref{fig:phase-shifts} we show the phase shifts for the
SI-corrected and uncorrected $f$-channels of Ce.
It can be seen that the
uncorrected $f$-states have a very sharp resonance just above the
Fermi energy.  The steep resonance corresponds to a long Wigner
delay-time and indicates that the state is already well localized. The
self-interaction corrected $f$-state is shifted down in energy by
about 9 eV, and becomes a bound state (it lies below the muffin-tin
zero).

In Fig. \ref{SIC-charge} we present the density of states and the
integrated DOS for the self-interaction corrected $f$-channel. It can
be seen that by integrating Eq.~(\ref{eq:nL-SIC}) up to the Fermi
energy, one collects slightly more than one electron. This is because
there is a small density of states in the vicinity of the Fermi level
(note the scale of the left-hand side axis in Fig. \ref{SIC-charge}), 
which is due to slight hybridization of the SIC
channel with the other $f$-channels whose resonances occur in the
vicinity of the Fermi level (see Fig.~\ref{fig:phase-shifts}). Some
contribution to this density of states might also come from 5$f$
states.  This is implied by the behavior of the phase shifts in
Fig.~\ref{fig:phase-shifts}.  The sharp jump by $\pi$ indicates that
4$f$ state is a bound state of the SIC potential, and the slow rise of
the phase shift above the Fermi energy can be associated with the
progression towards the 5$f$ state.  Figure~\ref{SIC-charge} also
shows that the integrated DOS at the energy where the phase shift goes
through $\pi$, i.e., at about -2.5 eV, is slightly less than 1.  This
is most likely due to the integration method used to display the
quantities in Fig. \ref{SIC-charge} which is less accurate than the
contour integral used in the self-consistent calculations.


\subsection{Total energies and equilibrium volumes of Ce $\alpha$ and $\gamma$ phases} 

%

\begin{figure}
\includegraphics[scale=0.7]{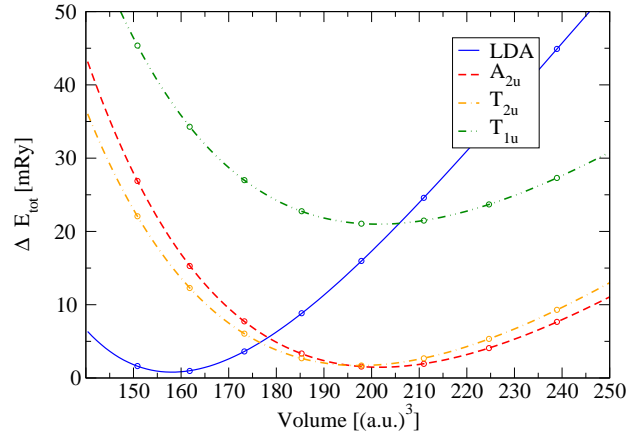}
\caption{\label{Ce-Etot} (color online) The calculated total energies for Ce from LDA and SIC-LSD,
 with different $f$-states localized, as functions of volume, given in atomic units (a.u.)$^3$.}
\end{figure}

In order to determine the ground state configuration of Ce at a given
volume, we calculated the total energies for different volumes using
the LDA to describe the $\alpha$-phase and the SIC-LSD formalism for 
the $\gamma$-phase, when
SI-correcting one localized $f$-electron, allowed to populate in sequence all 
possible $f$-states.  
In both LDA and SIC-LSD calculations spin-orbit coupling  has been
neglected for valence electrons, but fully included for core electrons,
for which the Dirac equation has been solved.
The corresponding total energies as functions of volume are shown in Fig.
\ref{Ce-Etot}. We find that the LDA, used to represent the $\alpha$-phase,
yields the lowest energy minimum, as seen in Table ~\ref{Ce-Etot-tab}. There
the ground state properties of the all studied configurations are summarized.
Table ~\ref{Ce-compare-tab} compares the present results for the ground state
configurations with previous calculations and with experimental values. 
Note small differences between the different calculations, which
are due to different schemes, and indicate the sensitivity
of the results to computational details. 

The observed degeneracy of the
states within the triplets demonstrates the rotational invariance of
the formalism.  Note the large crystal field (CF) splitting,
separating the $T_{1u}$ triplet from the other SIC-states. As already
mentioned, the calculations presented in this section assume a
ferromagnetic alignment of the local moments in the $\gamma$-phase.
However, when discussing the phase diagram of Ce, we will also consider the
disorder of the local moments using the DLM framework.\cite{DLM1,DLM2}

%

\begin{table}
\caption{\label{Ce-Etot-tab}
The total energy differences as obtained from the LDA and SIC-LSD calculations, with respect to the 
ground state energy solution (LDA), for Ce in different $f$-configurations. The corresponding 
volumes and bulk moduli (evaluated at the theoretical lattice constants) are also given.}
\begin{tabular}{cc|ccc}
\hline
\hline
& & $\Delta$E (mRy) & $V$ (a.u.)$^{3}$ & B (kbar) \\
\hline
LDA &     &  0.0 & 158 & 701 \\
    & A$_{2u}$ &  0.8 & 202 & 355 \\
SIC & T$_{1u}$ & 20.3 & 201 & 352 \\
    & T$_{2u}$ &  1.5 & 197 & 351 \\
\hline
\hline
\end{tabular}
\end{table}
\begin{table}
\caption{\label{Ce-compare-tab} Comparison of the computed equilibrium volumes and
bulk moduli with those of other calculations and experiment. The bulk moduli have been
calculated at the theoretical equilibrium volumes. Note that unlike in the present, 
SIC-LSD (KKR) implementation, the results based on the LMTO refer to the full SIC-LSD
scheme, involving repeated transformations between real and reciprocal
spaces, and Bloch and Wannier representations. The two different sets of LMTO calculations
refer to different basis sets and to different ways of solving the SIC-LSD eigenvalue problem.
In the reported LDA and GGA calculations the $\gamma$-phase was modelled by constraining the 
$f$-electrons to the core.}
\begin{tabular}{c|cccc}
\hline
\hline
Method             & \multicolumn{2}{c}{$\alpha$-Ce} & \multicolumn{2}{c}{$\gamma$-Ce}\\
                   & $V$ (\AA$^3$) & B (kbar)        & $V$ (\AA$^3$) & B (kbar) \\
\hline
SIC-LSD (KKR)$^a$  & 23.4 & 701 & 29.9 & 355 \\
SIC-LSD (LMTO)$^b$ & 24.7 & 484 & 32.6 & 310 \\
SIC-LSD (LMTO)$^c$ & 25.9 & 443 & 34.0 & 340 \\
LDA$^d$            & 24.5 & 477 & 33.7 & 312 \\
GGA$^d$            & 27.7 & 391 & 37.3 & 288 \\
Exp.$^e$           & 28.2 & 270 & 34.7 & 239 \\
\hline
\hline
\end{tabular}
\begin{flushleft}
$^a$ this work\\
$^b$ Ref. \onlinecite{SzotekEtAl:94} \\
$^c$ Ref. \onlinecite{Svane:96} \\
$^d$ Ref. \onlinecite{JohanssonEtAl:95} \\
$^e$ taken from Ref. \onlinecite{JohanssonEtAl:95}
\end{flushleft}
\end{table}

For the $\gamma-$phase, treated ferromagnetically, out of the three 
possible localized states listed in Table \ref{Ce-Etot-tab}, the state 
with the $A_{2u}$ symmetry gives the lowest energy solution.
This localized state is also
associated with the highest volume among the all possible 
localized configurations. Only 0.8 mRy separate the minima of the
$\alpha$ and $\gamma$ phases, giving rise to the transition
pressure at the absolute zero of about -2.3 kbar.  This is in
good agreement with the experimental value of -7 kbar, when
extrapolated to zero temperature, and with other theoretical values
(see also Table ~\ref{crit-points}).  The bulk moduli, given in Table \ref{Ce-Etot-tab},
are calculated at the theoretical equilibrium volumes. When evaluated at
the experimental volumes (as it is common practice in DFT calculations),
their values are substantially reduced to 239 kbar for the $\alpha$-phase and
203 kbar for the $\gamma$-phase, which is in considerably better agreement with 
the experimental numbers. The volume collapse (with respect to
the volume of the $\gamma$-phase) is obtained at 22\%, which also
compares well with the experimental values of 15\%-17\%.
We note that the underestimation of the volumes of both the $\alpha$-
and $\gamma$-phases is due to the KKR $l-$convergence problem,
which was addressed by Moghadam {\em et al.}\cite{KKR_l-conv}  They
demonstrated that angular momenta as high as 16 were needed to obtain
satisfactory convergence in the total energy.
In the present calculations we chose $l_{\rm max}=3$, which does not
seem sufficient for a good description of the equilibrium volumes of
the two phases. Although it seems that this $l-$convergence problem
should affect the LSD and SIC-LSD calculations in a similar manner, we
see a significantly larger error for the $\alpha$-phase, in agreement
with the results obtained by other well known  KKR
packages when the LDA approximation is implemented to describe the
electronic structure of the $\alpha-$phase.\cite{Julie:private} The
larger error for the $\alpha$-phase than for the $\gamma-$phase (found
also in the LMTO-ASA calculations) is most likely due to the fact that
LDA is not adequate for describing the experimentally reported correlated
nature of the $\alpha$-phase. In fact, the LDA calculations correspond strictly
to the high-pressure $\alpha'$-phase, which is purely tetravalent and
has a smaller lattice constant than the observed $\alpha$-phase.
However, as already mentioned, in our calculations we have treated Ce as 
a trivalent system (one localized $f$-electron) in the $\gamma$-phase,
and a tetravalent system (all $f$-electrons are treated as delocalized) in 
the $\alpha$-phase. Experimental data seems to suggest, that $\alpha$-Ce has 
a non-integer valence of 3.67. One could argue that this intermediate
valence character of the 4$f$ state could be represented in terms of a
pseudoalloy composed of the trivalent and tetravalent Ce atoms. We
shall elaborate on this point in one of the following subsections.

\subsection{Densities of states of Ce $\alpha$ and $\gamma$ phases} 
\label{subsec:a-g-DOS}
%

%
\begin{figure}
\includegraphics[scale=0.8]{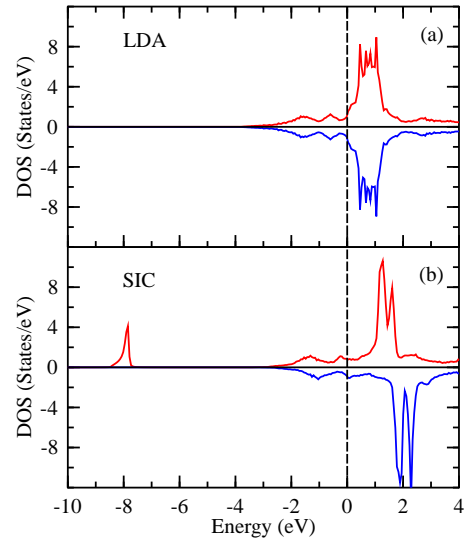}
\caption{\label{Ce-total-DOS} (color online)
Spin-resolved density of states of Ce in the $\alpha$- (a) and 
$\gamma$-phase (b) with ferromagnetic arrangement of local moments.}
\end{figure}
The densities of states of Ce from the LDA and ferromagnetic 
SIC-LSD calculations are shown in Fig. \ref{Ce-total-DOS}. 
The LDA DOS shows all the 
$f$-states hybridized into the $s$-, $p$- and $d$-states. 
However, in the SIC-LSD panel of the figure, one
clearly sees the split-off localized $f$-state at about -8 eV. Of course,
this does not agree with the spectroscopic position of this state. To
accomplish the latter, one would have to take into account the
self-energy, which could be evaluated from the total energy difference
corresponding to systems with constrained
$f$-occupations.\cite{TemmSzotek93}
The localized $f$-state apart, one can clearly see the exchange
splitting of the remaining states in the SIC-LSD calculations.
The unoccupied $f$-states in the SIC-LSD density of states are pushed up
by 1 eV or so. These unoccupied $f$-states are furthermore exchange split by 1 eV.

\begin{figure}
\includegraphics[scale=1.0]{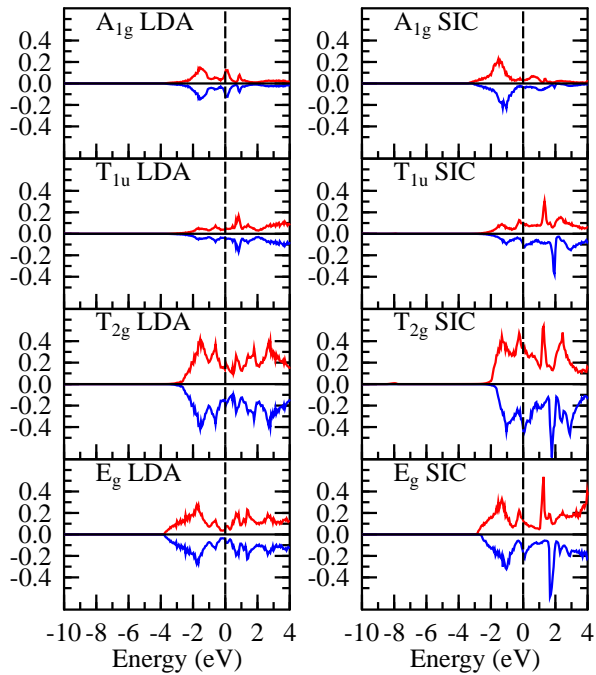}
\caption{
\label{spd-DOS} (color online)
Spin- and symmetry-resolved DOS (in states/eV) for the states originating 
from $s (A_{1g})$-, $p (T_{1u})$- 
and $d (T_{2g} \, \mbox{and} \, E_g)$- channels. As in Fig. \ref{Ce-total-DOS}, the SIC-LSD
calculation refers to the ferromagnetic arrangement of the local moments.}
\end{figure}

\begin{figure}
\includegraphics[scale=1.0]{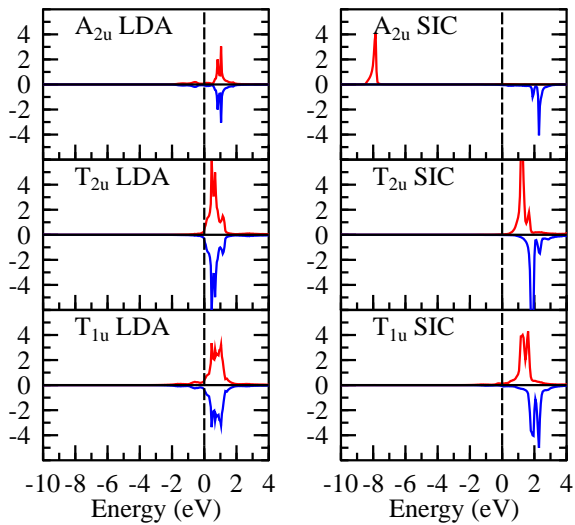}
\caption{
\label{f-DOS} (color online)
Spin- and symmetry-resolved DOS (in states/eV) for the states originating from $f$- channels. Note the
different scale of the plot with respect to Fig. \ref{spd-DOS}. As in Fig.~\ref{Ce-total-DOS}, 
the SIC-LSD calculation refers to the ferromagnetic arrangement of the local moments.}
\end{figure}

From Table \ref{Ce-charge-tab} we note that the number of occupied
$f$-states hardly differs between the $\gamma$-phase and the
$\alpha$-phase.  This non-integer value of 1.35 in both cases
is a consequence of the hybridization of the $f$-states with
the $s$, $p$ and $d$ states seen in Figs. \ref{spd-DOS} and
\ref{f-DOS}.  The number of $f$ electrons remains
constant between the LDA and the SIC-LSD, which might seem rather
surprising. What happens is that $f$ electrons that participate in
bonding in the $\alpha$-phase get transferred to localized
electrons in the $\gamma$-phase. At the same time some of the bonding
$d$-electrons are transferred to the repulsive $sp$-channel. These
effects conspire to give the larger lattice constant for the
$\gamma$-phase.


The $s$, $p$ and $d$ spin-resolved densities of states of Fig. \ref {spd-DOS} show
remarkably rigid band behavior between the LDA and SIC with the Fermi energy 
moving down with respect to the SIC-LSD $d$ partial density of states.
We also note that in Fig.~\ref{f-DOS} the
unoccupied $f$-states are well separated from the Fermi level, because
the Fermi energy is lowered, and the hybridization with the occupied
$s$, $p$ and $d$ states has been substantially reduced. These changes in
the $s$, $p$ and $d$ densities of states are also reflected in Table
\ref{Ce-charge-tab} where we see a reduction of 0.2 electrons in the
$d$-channel of the localized phase and a corresponding increase of 0.1
electrons in both the $s$ and $p$ channels of the localized phase.
Even though the Fermi energy moves down in the localized phase, we see
from Fig.~\ref {spd-DOS} that the number of occupied states in the
$s$ and $p$ channels has increased.

\begin{table}
\caption{\label{Ce-charge-tab}Angular momentum decomposed charges from LDA and SIC-LSD calculations.
Note that the p-channel includes the 5p semi-core states.}
\begin{tabular}{c|cccc}
\hline \hline
 & $s$ & $p$ & $d$ & $f$ \\
\hline
LDA & \: 0.40 \: & \: 6.06 \: & \: 2.19 \: & \: 1.35 \: \\
SIC & 0.51 & 6.16 & 1.99  & 1.35 \\
\hline \hline
\end{tabular}
\vspace{4mm}

\end{table}

\subsection{Ce $\alpha - \gamma$ pseudoalloy}

In order to improve on the LDA representation of correlations in the
$\alpha$-phase of Ce, in the present approach, in the spirit of the 
Hubbard III approximation,\cite{Hubbard3} one can model the
experimentally implied non-integer valence of the Ce ions by a pseudoalloy
consisting of the trivalent (SIC-LSD) Ce ions with concentration c,
and the tetravalent (LDA) Ce ions with the concentration (1-c).
In addition, taking into account the disordered local moments
of the trivalent Ce ions in the $\gamma$-phase, one can assume that
their up and down orientations occur with equal probabilities.
Supposing homogeneous randomness, such a ternary
pseudoalloy can be described by the coherent potential approximation
(CPA). The respective concentrations of the trivalent and 
tetravalent Ce ions in the pseudoalloy are then determined by
minimizing the total energy for each volume with respect to the
concentration $c$.

\begin{figure}
\begin{picture}(0,0)%
\includegraphics{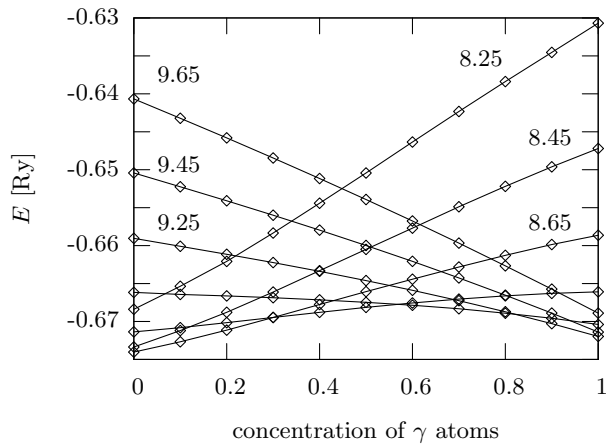}%
\end{picture}%
\begingroup
\setlength{\unitlength}{0.0200bp}%
\begin{picture}(12599,8640)(0,0)%
\put(2750,2366){\makebox(0,0)[r]{\strut{}-0.67}}%
\put(2750,3797){\makebox(0,0)[r]{\strut{}-0.66}}%
\put(2750,5228){\makebox(0,0)[r]{\strut{}-0.65}}%
\put(2750,6659){\makebox(0,0)[r]{\strut{}-0.64}}%
\put(2750,8090){\makebox(0,0)[r]{\strut{}-0.63}}%
\put(3025,1100){\makebox(0,0){\strut{} 0}}%
\put(4775,1100){\makebox(0,0){\strut{} 0.2}}%
\put(6525,1100){\makebox(0,0){\strut{} 0.4}}%
\put(8275,1100){\makebox(0,0){\strut{} 0.6}}%
\put(10025,1100){\makebox(0,0){\strut{} 0.8}}%
\put(11775,1100){\makebox(0,0){\strut{} 1}}%
\put(9500,7300){\makebox(0,0){\strut{} 8.25}}
\put(10800,5800){\makebox(0,0){\strut{} 8.45}}
\put(10800,4200){\makebox(0,0){\strut{} 8.65}}
\put(3800,7000){\makebox(0,0){\strut{} 9.65}}
\put(3800,5300){\makebox(0,0){\strut{} 9.45}}
\put(3800,4200){\makebox(0,0){\strut{} 9.25}}
\put(950,4870){\rotatebox{90}{\makebox(0,0){\strut{}$E$ [Ry]}}}%
\put(7400,275){\makebox(0,0){\strut{}concentration of $\gamma$ atoms}}%
\end{picture}%
\endgroup

\caption{
\label{CPA-fig}
Total energies (T=0K) of the Ce $\alpha-\gamma$ pseudoalloy as a function
of the concentration of localized states. The curves correspond to the lattice
constants, indicated in the figure. The labels of 8.85 and 9.05, corresponding to
the remaining curves, have been omitted for readability.
}
\end{figure}

In Fig. \ref{CPA-fig} we show the total energies for the
$\alpha$-$\gamma$ pseudoalloy at T=0K, in which the $\gamma$-phase
occurs with the concentration c (c/2 for each spin orientation), and 
the $\alpha$-phase with the concentration (1-c), for several lattice
constants. It can be seen that all shown total energy curves have
their minima either at c=0 (pure $\alpha$-phase) or c=1
(pure $\gamma$-phase).  Hence a fractional occupation of the 4$f$ state
appears to be energetically unfavorable for all lattice constants.
From these calculations we can conclude that a static,
single-site approximation is not sufficient to describe the
intermediate valence state of $\alpha$-Ce at T=0 K. These calculations
are consistent with the earlier results by Svane who performed supercell
calculations to model 25\%, 50\% and 75\% of $\alpha$-$\gamma$
admixtures, but treating the $\gamma$-phase ferromagnetically
and not as a DLM phase.\cite{Svane:96} There too, no total energy minimum was found for
intermediate concentrations between 0 and 1, and also a mainly convex (from
above) curvature for the total energy, as a function of concentration,
was obtained. This suggests that to describe the intermediate valence
state of the $\alpha$-phase one would need to consider
a dynamical generalization of the CPA,\cite{Kakehashi} which would involve dynamical
fluctuations between the trivalent and tetravalent states.
Other possible mechanisms to favor intermediate valence will be commented
about in section \ref{sec:discussion}.

\subsection{Ce $\alpha - \gamma$ spectral functions} 

\begin{figure}
\vspace*{-1.0cm}
\begin{picture}(0,0)%
\includegraphics{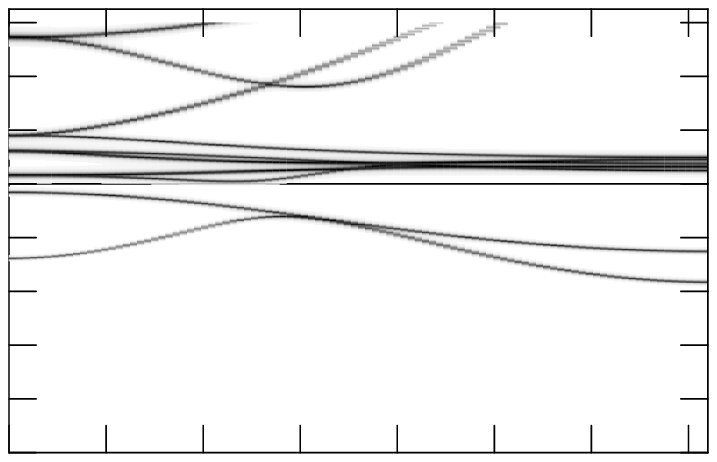}%
\end{picture}%
\begingroup
\setlength{\unitlength}{0.0200bp}%
\begin{picture}(14400,10800)(0,0)%
\put(6775,942){\makebox(0,0){\strut{}$k_x$}}%
\put(424,5675){\rotatebox{90}{\makebox(0,0){\strut{}$\epsilon$ [eV]}}}%
\put(1743,1767){\makebox(0,0){\strut{} 0}}%
\put(3141,1767){\makebox(0,0){\strut{} 0.1}}%
\put(4539,1767){\makebox(0,0){\strut{} 0.2}}%
\put(5937,1767){\makebox(0,0){\strut{} 0.3}}%
\put(7334,1767){\makebox(0,0){\strut{} 0.4}}%
\put(8732,1767){\makebox(0,0){\strut{} 0.5}}%
\put(10129,1767){\makebox(0,0){\strut{} 0.6}}%
\put(11527,1767){\makebox(0,0){\strut{} 0.7}}%
\put(1386,2482){\makebox(0,0)[r]{\strut{}-10}}%
\put(1386,3256){\makebox(0,0)[r]{\strut{}-8}}%
\put(1386,4030){\makebox(0,0)[r]{\strut{}-6}}%
\put(1386,4805){\makebox(0,0)[r]{\strut{}-4}}%
\put(1386,5579){\makebox(0,0)[r]{\strut{}-2}}%
\put(1386,6352){\makebox(0,0)[r]{\strut{} 0}}%
\put(1386,7126){\makebox(0,0)[r]{\strut{} 2}}%
\put(1386,7900){\makebox(0,0)[r]{\strut{} 4}}%
\put(1386,8674){\makebox(0,0)[r]{\strut{} 6}}%
\put(11000,8200){\makebox(0,0)[r]{\strut{} (a)}}%
\end{picture}%
\endgroup\\
\vspace*{-1.5cm}
\begin{picture}(0,0)%
\includegraphics{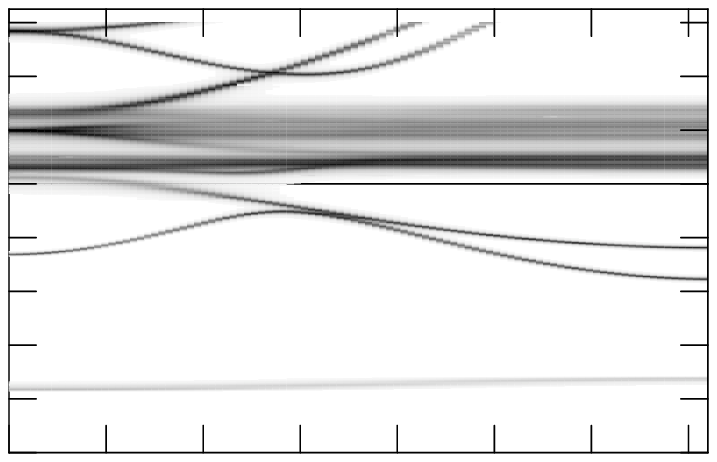}%
\end{picture}%
\begingroup
\setlength{\unitlength}{0.0200bp}%
\begin{picture}(14400,10800)(0,0)%
\put(6775,942){\makebox(0,0){\strut{}$k_x$}}%
\put(424,5675){\rotatebox{90}{\makebox(0,0){\strut{}$\epsilon$ [eV]}}}%
\put(1743,1767){\makebox(0,0){\strut{} 0}}%
\put(3141,1767){\makebox(0,0){\strut{} 0.1}}%
\put(4539,1767){\makebox(0,0){\strut{} 0.2}}%
\put(5937,1767){\makebox(0,0){\strut{} 0.3}}%
\put(7334,1767){\makebox(0,0){\strut{} 0.4}}%
\put(8732,1767){\makebox(0,0){\strut{} 0.5}}%
\put(10129,1767){\makebox(0,0){\strut{} 0.6}}%
\put(11527,1767){\makebox(0,0){\strut{} 0.7}}%
\put(1386,2482){\makebox(0,0)[r]{\strut{}-10}}%
\put(1386,3256){\makebox(0,0)[r]{\strut{}-8}}%
\put(1386,4030){\makebox(0,0)[r]{\strut{}-6}}%
\put(1386,4805){\makebox(0,0)[r]{\strut{}-4}}%
\put(1386,5579){\makebox(0,0)[r]{\strut{}-2}}%
\put(1386,6352){\makebox(0,0)[r]{\strut{} 0}}%
\put(1386,7126){\makebox(0,0)[r]{\strut{} 2}}%
\put(1386,7900){\makebox(0,0)[r]{\strut{} 4}}%
\put(1386,8674){\makebox(0,0)[r]{\strut{} 6}}%
\put(11000,8200){\makebox(0,0)[r]{\strut{} (b)}}%
\end{picture}%
\endgroup\\
\vspace{-1.5cm}
\begin{picture}(0,0)%
\includegraphics{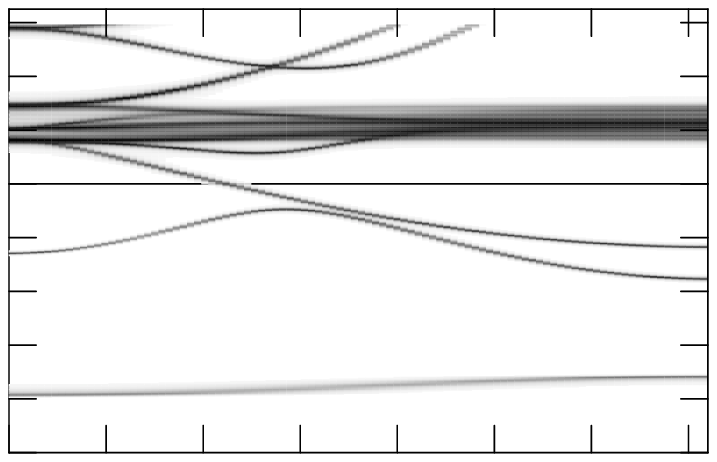}%
\end{picture}%
\begingroup
\setlength{\unitlength}{0.0200bp}%
\begin{picture}(14400,10800)(0,0)%
\put(6775,942){\makebox(0,0){\strut{}$k_x$}}%
\put(424,5675){\rotatebox{90}{\makebox(0,0){\strut{}$\epsilon$ [eV]}}}%
\put(1743,1767){\makebox(0,0){\strut{} 0}}%
\put(3141,1767){\makebox(0,0){\strut{} 0.1}}%
\put(4539,1767){\makebox(0,0){\strut{} 0.2}}%
\put(5937,1767){\makebox(0,0){\strut{} 0.3}}%
\put(7334,1767){\makebox(0,0){\strut{} 0.4}}%
\put(8732,1767){\makebox(0,0){\strut{} 0.5}}%
\put(10129,1767){\makebox(0,0){\strut{} 0.6}}%
\put(11527,1767){\makebox(0,0){\strut{} 0.7}}%
\put(1386,2482){\makebox(0,0)[r]{\strut{}-10}}%
\put(1386,3256){\makebox(0,0)[r]{\strut{}-8}}%
\put(1386,4030){\makebox(0,0)[r]{\strut{}-6}}%
\put(1386,4805){\makebox(0,0)[r]{\strut{}-4}}%
\put(1386,5579){\makebox(0,0)[r]{\strut{}-2}}%
\put(1386,6352){\makebox(0,0)[r]{\strut{} 0}}%
\put(1386,7126){\makebox(0,0)[r]{\strut{} 2}}%
\put(1386,7900){\makebox(0,0)[r]{\strut{} 4}}%
\put(1386,8674){\makebox(0,0)[r]{\strut{} 6}}%
\put(11000,8200){\makebox(0,0)[r]{\strut{} (c)}}%
\end{picture}%
\endgroup\\[-0.5cm]
\caption{
\label{spectral-function-GX}
Spectral functions along $\Gamma-X$, relative to the Fermi level,
for the $\gamma$-phase concentration 0 in panel (a), 0.5 in (b) and 1 in (c), respectively.
The spectral functions were calculated at the lattice constant $a=8.65 \, a.u.$}
\end{figure}

In this section we discuss the spectral functions along the $\Gamma-X$ line, calculated for the
pseudoalloy consisting of 50\% $\gamma$-admixture into the $\alpha$-phase, 
i.e., 50\% $\alpha$, 25\% $\gamma$ spin-up and 25 \% $\gamma$ spin-down, in
comparison with the spectral functions of the pure phases, all at the same volume,
as shown in Fig. \ref{spectral-function-GX}. The pure $\gamma$-phase has been represented 
by a 50 \% spin-up and 50 \% spin-down alloy.
The pure $\alpha$-phase (panel
(a)) shows a well defined band-structure. (The minor smearing of the bands is due
to a small imaginary part added to the energy).
Of course, the LDA leads trivially to a non spin-polarized band structure, but
the absence of an exchange splitting in the panels (b) and (c) is
due to the use of DLM, which defines an effective medium in which
the local moments are averaged out. In the panels (b) and (c) 
the broadening of the spectral functions is apparent. 
\begin{figure}
\begin{picture}(0,0)%
\includegraphics{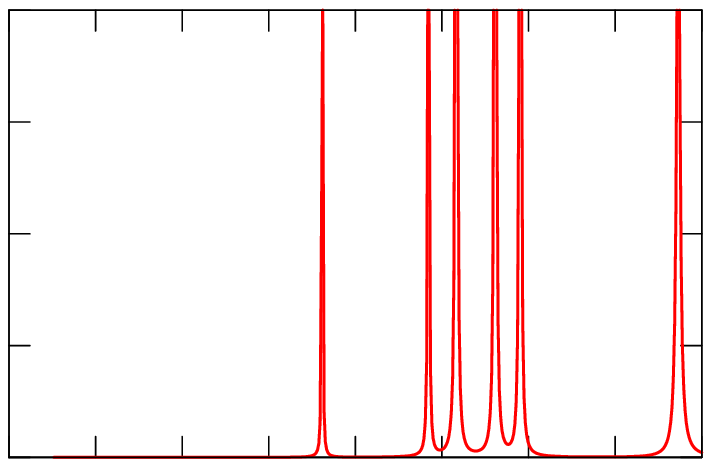}%
\end{picture}%
\begingroup
\setlength{\unitlength}{0.0200bp}%
\begin{picture}(10800,8640)(0,0)%
\put(-275,1650){\makebox(0,0)[r]{\strut{} 0}}%
\put(-275,3260){\makebox(0,0)[r]{\strut{} 5}}%
\put(-275,4870){\makebox(0,0)[r]{\strut{} 10}}%
\put(-275,6480){\makebox(0,0)[r]{\strut{} 15}}%
\put(-275,8090){\makebox(0,0)[r]{\strut{} 20}}%
\put(0,1100){\makebox(0,0){\strut{}-10}}%
\put(1247,1100){\makebox(0,0){\strut{}-8}}%
\put(2494,1100){\makebox(0,0){\strut{}-6}}%
\put(3741,1100){\makebox(0,0){\strut{}-4}}%
\put(4988,1100){\makebox(0,0){\strut{}-2}}%
\put(6234,1100){\makebox(0,0){\strut{} 0}}%
\put(7481,1100){\makebox(0,0){\strut{} 2}}%
\put(8728,1100){\makebox(0,0){\strut{} 4}}%
\put(9975,1100){\makebox(0,0){\strut{} 6}}%
\put(-1050,4870){\rotatebox{90}{\makebox(0,0){\strut{}A$_\Gamma(\epsilon)$}}}%
\put(4987,475){\makebox(0,0){\strut{}$\epsilon$ [eV]}}%
\put(4400,8400){\makebox(0,0){\strut{} s}}
\put(5900,8400){\makebox(0,0){\strut{} f}}
\put(6400,8400){\makebox(0,0){\strut{} f}}
\put(6900,8400){\makebox(0,0){\strut{} f}}
\put(7280,8400){\makebox(0,0){\strut{} d}}
\put(9600,8400){\makebox(0,0){\strut{} d}}
\put(1000,7200){\makebox(0,0){\strut{} (a)}}
\end{picture}%
\endgroup\\
\begin{picture}(0,0)%
\includegraphics{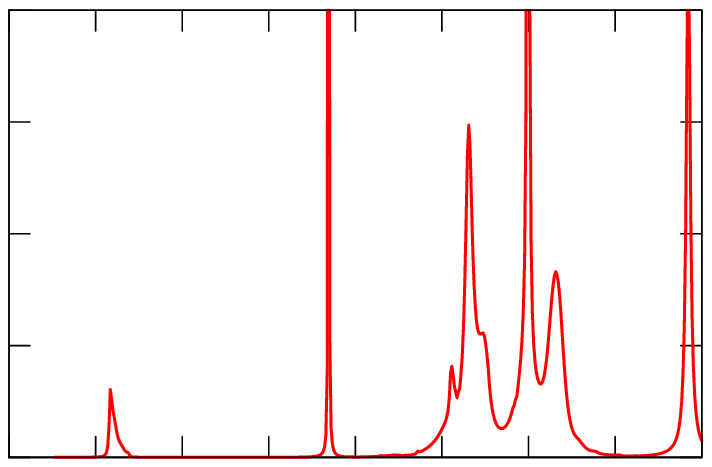}%
\end{picture}%
\begingroup
\setlength{\unitlength}{0.0200bp}%
\begin{picture}(10800,8640)(0,0)%
\put(-275,1650){\makebox(0,0)[r]{\strut{} 0}}%
\put(-275,3260){\makebox(0,0)[r]{\strut{} 5}}%
\put(-275,4870){\makebox(0,0)[r]{\strut{} 10}}%
\put(-275,6480){\makebox(0,0)[r]{\strut{} 15}}%
\put(-275,8090){\makebox(0,0)[r]{\strut{} 20}}%
\put(0,1100){\makebox(0,0){\strut{}-10}}%
\put(1247,1100){\makebox(0,0){\strut{}-8}}%
\put(2494,1100){\makebox(0,0){\strut{}-6}}%
\put(3741,1100){\makebox(0,0){\strut{}-4}}%
\put(4988,1100){\makebox(0,0){\strut{}-2}}%
\put(6234,1100){\makebox(0,0){\strut{} 0}}%
\put(7481,1100){\makebox(0,0){\strut{} 2}}%
\put(8728,1100){\makebox(0,0){\strut{} 4}}%
\put(9975,1100){\makebox(0,0){\strut{} 6}}%
\put(-1050,4870){\rotatebox{90}{\makebox(0,0){\strut{}A$_\Gamma(\epsilon)$}}}%
\put(4987,475){\makebox(0,0){\strut{}$\epsilon$ [eV]}}%
\put(1300,8400){\makebox(0,0){\strut{} f}}
\put(4400,8400){\makebox(0,0){\strut{} s}}
\put(5900,8400){\makebox(0,0){\strut{} f}}
\put(6400,8400){\makebox(0,0){\strut{} f}}
\put(7280,8400){\makebox(0,0){\strut{} d}}
\put(7900,8400){\makebox(0,0){\strut{} f}}
\put(9600,8400){\makebox(0,0){\strut{} d}}
\put(1000,7200){\makebox(0,0){\strut{} (b)}}
\end{picture}%
\endgroup\\
\begin{picture}(0,0)%
\includegraphics{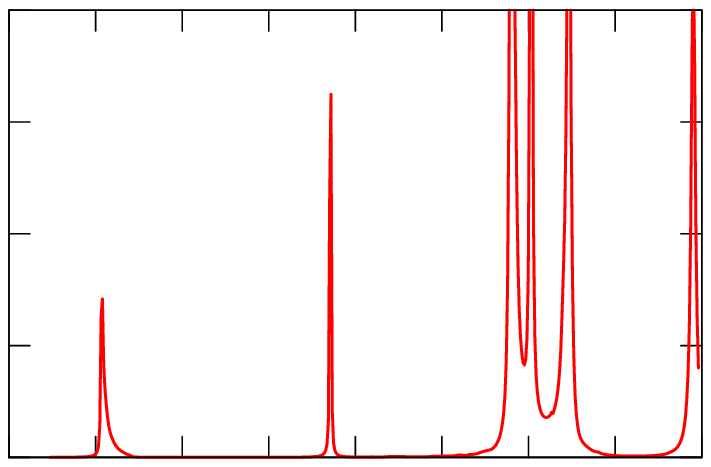}%
\end{picture}%
\begingroup
\setlength{\unitlength}{0.0200bp}%
\begin{picture}(10800,8640)(0,0)%
\put(-275,1650){\makebox(0,0)[r]{\strut{} 0}}%
\put(-275,3260){\makebox(0,0)[r]{\strut{} 5}}%
\put(-275,4870){\makebox(0,0)[r]{\strut{} 10}}%
\put(-275,6480){\makebox(0,0)[r]{\strut{} 15}}%
\put(-275,8090){\makebox(0,0)[r]{\strut{} 20}}%
\put(0,1100){\makebox(0,0){\strut{}-10}}%
\put(1247,1100){\makebox(0,0){\strut{}-8}}%
\put(2494,1100){\makebox(0,0){\strut{}-6}}%
\put(3741,1100){\makebox(0,0){\strut{}-4}}%
\put(4988,1100){\makebox(0,0){\strut{}-2}}%
\put(6234,1100){\makebox(0,0){\strut{} 0}}%
\put(7481,1100){\makebox(0,0){\strut{} 2}}%
\put(8728,1100){\makebox(0,0){\strut{} 4}}%
\put(9975,1100){\makebox(0,0){\strut{} 6}}%
\put(-1050,4870){\rotatebox{90}{\makebox(0,0){\strut{}A$_\Gamma(\epsilon)$}}}%
\put(4987,475){\makebox(0,0){\strut{}$\epsilon$ [eV]}}%
\put(1300,8400){\makebox(0,0){\strut{} f}}
\put(4400,8400){\makebox(0,0){\strut{} s}}
\put(7000,8400){\makebox(0,0){\strut{} f}}
\put(7600,8400){\makebox(0,0){\strut{} d}}
\put(8000,8400){\makebox(0,0){\strut{} f}}
\put(9600,8400){\makebox(0,0){\strut{} d}}
\put(1000,7200){\makebox(0,0){\strut{} (c)}}
\end{picture}%
\endgroup\\
\caption{
\label{spectral-function-G} (color online)
Spectral functions at the $\Gamma$ point for the $\gamma$-phase concentrations 0 (a), 0.5 (b)
and 1 (c). The character of the respective peaks is marked above the upper horizontal line 
of each panel.}
\end{figure}
The actual line-width of the spectral function can clearly be seen in
Fig. \ref{spectral-function-G}, showing the spectral functions at
the $\Gamma$ point. Here, similarly to the smearing effect seen in 
Fig. \ref{spectral-function-GX},
the residual linewidth seen in panel (a) is purely
due to the small imaginary part of the energy, necessary to obtain well
behaved $\tau$-matrices.

The spectral functions reveal some features of the
current approach. Firstly, it can be seen that the $s$- and
$d$-derived states at the $\Gamma$-point are hardly affected by the
CPA. These states have no hybridization with the SI-corrected $f$-state.
Symmetry analysis of the spectral function, shown in panel (b) of 
Fig. \ref{spectral-function-G}, reveals that
the $A_{2u}$ $f$-state appears twice (the peak at -8 eV and
the sharp shoulder just above the Fermi energy), since we are in the
split-band regime.  Due to the DLM treatment of the $\gamma$-phase,
this feature is also seen in panel (c), where the upper of the
split-band peaks merges with the lower triplet $f$-peak. The two
triplets ($T_{1u}$ and $T_{2u}$) show common band behavior.
The corresponding broadenings are very different between panels (b)
and (c).  Moreover, panel (c), as compared to panel
(a), also shows that the unoccupied $f$-states have been pushed up in energy,
because the localized $f$-electron is more effective in screening
the nuclear charge.
This results in an energy splitting of the $\alpha$ and $\gamma$
unoccupied triplets which leads to a broadening of the
unoccupied triplets, as seen in panel (b). The shoulder at
about 1 eV is an indication that the splitting is noticeable on the
scale of the dispersion of the bands. The broadening of the triplets
is reduced in panel (c) where no $\alpha$-phase is admixed and where
the broadening of the lower triplet state is a consequence of 
merging with the upper split-band peak of the singlet state.
Note that the $f$-states at the Fermi-level have a
finite lifetime which might indicate a shortcoming of the static
CPA for the description of an intermediate valence: a coherent mixture
of the localized and delocalized states in terms of wave functions
might be a more appropriate description.

\subsection{Phase diagram}
\label{sec:phase-diag}

In this subsection we concentrate on
the finite temperature phase diagram of Ce.
The idea of describing Ce at finite temperatures as a pseudoalloy of
$\alpha$- and $\gamma$-Ce atoms was first put forward by Johansson {\em et
  al.} \cite{JohanssonEtAl:95} and by Svane.\cite{Svane:96}  
Since at finite temperatures the thermal (classical) fluctuations
are of major importance, the static approximation should suffice. In the
work by Johansson {\em et al.} the pseudoalloy was treated by the CPA 
implemented within the LMTO method,
where the $\gamma$-phase was modeled by including the 4$f$-states into
the core, while in the $\alpha$-phase the $f$ states were treated as
band-states.  Due to the different treatment of both phases, their total
energies could not be compared and the energies of the
$\gamma$-phase had to be adjusted by hand to yield the correct 
$\alpha-\gamma$ transition pressure at zero temperature.
Svane, on the other hand, described the $\gamma$-phase as a
ferromagnet using the LMTO-SIC, thus treating both phases on equal
footing, and utilizing a supercell to mimic the pseudoalloy at only a few
accessible concentrations. From these calculations he concluded that a
linear interpolation of the $\alpha$- and $\gamma$-energies to
arbitrary concentrations should be adequate enough.

Here we present calculations which combine both the CPA
and the SIC-LSD to describe the $\gamma$-phase as a DLM system,
treated as a ternary alloy, consisting of spin-up and spin-down SIC
sites with concentrations $c/2$ each, and LDA sites with the concentration
$(1-c)$.
In addition,  we go beyond the scope of previous works by taking into
account the effect of finite temperatures on the electronic total
energies and the electronic contribution to the entropy, as defined
in section \ref{sec:FT}. However, the vibrational entropy, S$_{\rm vib}$, 
is neglected in the presented results. We shall
briefly comment on its effect on the calculated phase diagram in the next 
subsection where we analyze in detail how different aspects of the present
calculations influence the final results.

%
%

Ideally, one would like to treat a pseudoalloy, which consists of all
possible states of a Ce-ion, i.e. the LDA, and all possible SI-corrected
$f$ states. This
would give rise to a pseudoalloy consisting of 15 components: the LDA
plus the 14 possible $f$-SIC states (including the spin multiplicity).
Since this would be a quite formidable task, we use a simplified
approach. Figure~\ref{Ce-Etot} indicates that crystal-field
splitting gives rise to nearly degenerate $A_{2u}$ (singlet) and
$T_{2u}$ (triplet) states, while the $T_{1u}$ triplet lies 20 mRy
higher in energy. At the temperatures considered here, the $T_{1u}$ states are
thermally not accessible.  Thus treating the remaining eight states as
degenerate, leaves us with a nine component pseudoalloy, with the
constraint that the concentrations of the considered eight SIC states are 
equal and can be set to $c/8$. It is easy to show that in this case, in addition
to the mixing entropy defined in section \ref{sec:FT} one has to take into
account a term 
\begin{equation}
\label{eq:5}
S_{\rm mag}(c) = k_B c \ln 8 ,
\end{equation}
arising from these eight-fold multiplicity. Note that in previous studies, where the CF
splitting has not been taken into account, the magnetic entropy was
assumed to be that of a spin-orbit (SO) coupled $J=5/2$ state, i.e.
$S_{\rm mag}(c) = k_B c \ln 6$. In the next subsection we shall comment on how the
two different magnetic entropy terms influence the critical characteristics of the
calculated phase diagram.

\begin{figure}
\hspace*{-8cm}\begin{minipage}{7cm}
\hspace*{6cm}$T$=0K\\[-1.2cm]
\begin{picture}(0,0)%
\includegraphics{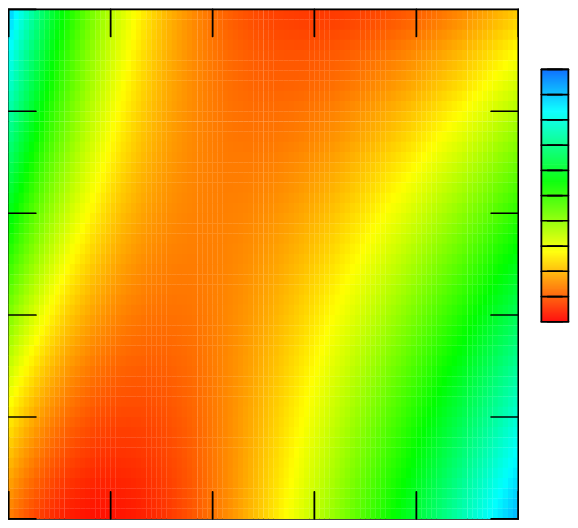}%
\end{picture}%
\begingroup
\setlength{\unitlength}{0.0200bp}%
\begin{picture}(19800,11880)(0,0)%
\put(14541,5360){\makebox(0,0)[l]{\strut{}-0.675}}%
\put(14541,5723){\makebox(0,0)[l]{\strut{}-0.67}}%
\put(14541,6087){\makebox(0,0)[l]{\strut{}-0.665}}%
\put(14541,6450){\makebox(0,0)[l]{\strut{}-0.66}}%
\put(14541,6814){\makebox(0,0)[l]{\strut{}-0.655}}%
\put(14541,7177){\makebox(0,0)[l]{\strut{}-0.65}}%
\put(14541,7541){\makebox(0,0)[l]{\strut{}-0.645}}%
\put(14541,7904){\makebox(0,0)[l]{\strut{}-0.64}}%
\put(14541,8268){\makebox(0,0)[l]{\strut{}-0.635}}%
\put(14541,8631){\makebox(0,0)[l]{\strut{}-0.63}}%
\put(14541,8995){\makebox(0,0)[l]{\strut{}-0.625}}%
\put(9900,1122){\makebox(0,0){\strut{}$V$ [a.u.$^3$]}}%
\put(4782,6190){\rotatebox{90}{\makebox(0,0){\strut{}$c$}}}%
\put(6232,1872){\makebox(0,0){\strut{} 140}}%
\put(7699,1872){\makebox(0,0){\strut{} 160}}%
\put(9167,1872){\makebox(0,0){\strut{} 180}}%
\put(10633,1872){\makebox(0,0){\strut{} 200}}%
\put(12101,1872){\makebox(0,0){\strut{} 220}}%
\put(13568,1872){\makebox(0,0){\strut{} 240}}%
\put(5907,2522){\makebox(0,0)[r]{\strut{} 0}}%
\put(5907,3989){\makebox(0,0)[r]{\strut{} 0.2}}%
\put(5907,5457){\makebox(0,0)[r]{\strut{} 0.4}}%
\put(5907,6923){\makebox(0,0)[r]{\strut{} 0.6}}%
\put(5907,8391){\makebox(0,0)[r]{\strut{} 0.8}}%
\put(5907,9858){\makebox(0,0)[r]{\strut{} 1}}%
\end{picture}%
\endgroup
\\[-0.6cm]
\hspace*{6cm}$T$=800K\\[-1.2cm]
\begin{picture}(0,0)%
\includegraphics{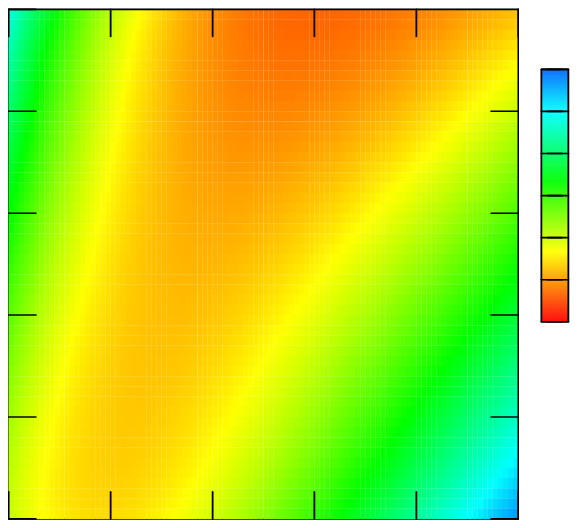}%
\end{picture}%
\begingroup
\setlength{\unitlength}{0.0200bp}%
\begin{picture}(19800,11880)(0,0)%
\put(14541,5360){\makebox(0,0)[l]{\strut{}-0.69}}%
\put(14541,5965){\makebox(0,0)[l]{\strut{}-0.68}}%
\put(14541,6571){\makebox(0,0)[l]{\strut{}-0.67}}%
\put(14541,7177){\makebox(0,0)[l]{\strut{}-0.66}}%
\put(14541,7783){\makebox(0,0)[l]{\strut{}-0.65}}%
\put(14541,8389){\makebox(0,0)[l]{\strut{}-0.64}}%
\put(14541,8995){\makebox(0,0)[l]{\strut{}-0.63}}%
\put(9900,1122){\makebox(0,0){\strut{}$V$ [a.u.$^3$]}}%
\put(4782,6190){\rotatebox{90}{\makebox(0,0){\strut{}$c$}}}%
\put(6232,1872){\makebox(0,0){\strut{} 140}}%
\put(7699,1872){\makebox(0,0){\strut{} 160}}%
\put(9167,1872){\makebox(0,0){\strut{} 180}}%
\put(10633,1872){\makebox(0,0){\strut{} 200}}%
\put(12101,1872){\makebox(0,0){\strut{} 220}}%
\put(13568,1872){\makebox(0,0){\strut{} 240}}%
\put(5907,2522){\makebox(0,0)[r]{\strut{} 0}}%
\put(5907,3989){\makebox(0,0)[r]{\strut{} 0.2}}%
\put(5907,5457){\makebox(0,0)[r]{\strut{} 0.4}}%
\put(5907,6923){\makebox(0,0)[r]{\strut{} 0.6}}%
\put(5907,8391){\makebox(0,0)[r]{\strut{} 0.8}}%
\put(5907,9858){\makebox(0,0)[r]{\strut{} 1}}%
\end{picture}%
\endgroup
\\[-0.6cm]
\hspace*{6cm}$T$=1600K\\[-1.2cm]
\begin{picture}(0,0)%
\includegraphics{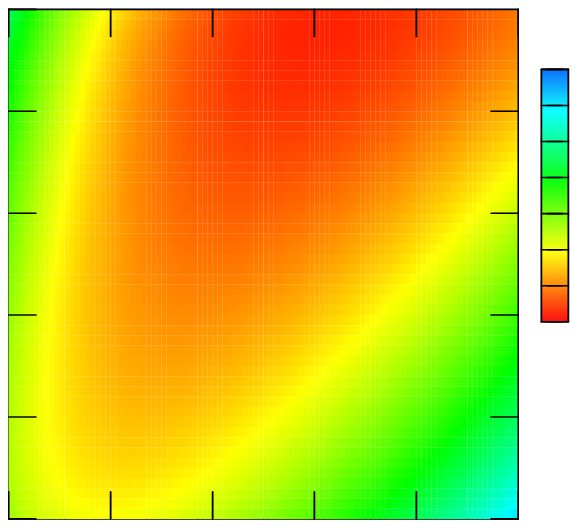}%
\end{picture}%
\begingroup
\setlength{\unitlength}{0.0200bp}%
\begin{picture}(19800,11880)(0,0)%
\put(14541,5360){\makebox(0,0)[l]{\strut{}-0.7}}%
\put(14541,5879){\makebox(0,0)[l]{\strut{}-0.69}}%
\put(14541,6398){\makebox(0,0)[l]{\strut{}-0.68}}%
\put(14541,6917){\makebox(0,0)[l]{\strut{}-0.67}}%
\put(14541,7437){\makebox(0,0)[l]{\strut{}-0.66}}%
\put(14541,7956){\makebox(0,0)[l]{\strut{}-0.65}}%
\put(14541,8475){\makebox(0,0)[l]{\strut{}-0.64}}%
\put(14541,8995){\makebox(0,0)[l]{\strut{}-0.63}}%
\put(9900,1122){\makebox(0,0){\strut{}$V$ [a.u.$^3$]}}%
\put(4782,6190){\rotatebox{90}{\makebox(0,0){\strut{}$c$}}}%
\put(6232,1872){\makebox(0,0){\strut{} 140}}%
\put(7699,1872){\makebox(0,0){\strut{} 160}}%
\put(9167,1872){\makebox(0,0){\strut{} 180}}%
\put(10633,1872){\makebox(0,0){\strut{} 200}}%
\put(12101,1872){\makebox(0,0){\strut{} 220}}%
\put(13568,1872){\makebox(0,0){\strut{} 240}}%
\put(5907,2522){\makebox(0,0)[r]{\strut{} 0}}%
\put(5907,3989){\makebox(0,0)[r]{\strut{} 0.2}}%
\put(5907,5457){\makebox(0,0)[r]{\strut{} 0.4}}%
\put(5907,6923){\makebox(0,0)[r]{\strut{} 0.6}}%
\put(5907,8391){\makebox(0,0)[r]{\strut{} 0.8}}%
\put(5907,9858){\makebox(0,0)[r]{\strut{} 1}}%
\end{picture}%
\endgroup
\\[-1.2cm]
\end{minipage}
\caption{
\label{free-energy-plots} (color) 
Calculated free energies for the temperatures $T$=0, 800 and 1600K.
The plots represent fits to the calculations, which have been performed
for concentrations from 0 (corresponding to the pure $\alpha$-phase) to 1 
(corresponding to the pure $\gamma$-phase), in steps of 0.1, and for lattice 
constants from 8.25 a.u. to 9.65 a.u., in steps of 0.2 a.u.
A constant of 17717 Ry has been added to all energies.
}
\end{figure}

We performed calculations for several lattice constants, embracing the
equilibrium lattice constants of both phases, and concentrations from
0 to 1, in steps of 0.1. The results for the free energies of the three
selected temperatures are shown in Fig. \ref{free-energy-plots}.
In the $T=0$K panel one clearly sees the two minima, corresponding to
the pure $\alpha-$phase (LDA) and the pure $\gamma-$phase (SIC-LSD)
calculations. The equilibrium lattice constant for a given
concentration $c$ interpolates between the two extremes, and it is
apparent that an intermediate valence state, even if not energetically
favorable, would correct the underestimated lattice constant of the
$\alpha$-phase.
As the temperature is increased, the free energy surface gets strongly
tilted towards the SIC-LSD side and now shows only one broad minimum. 
This is mainly due to the magnetic entropy.

\begin{figure}
\hspace*{-9cm}\begin{minipage}{7cm}
\hspace*{6cm}$T$=0K\\[-1.2cm]
\begin{picture}(0,0)%
\includegraphics{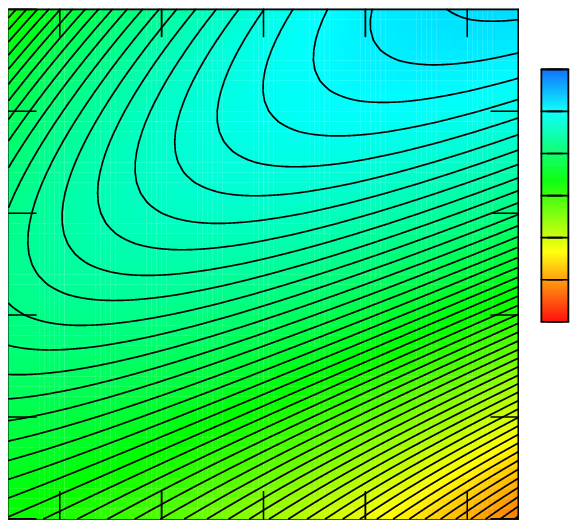}%
\end{picture}%
\begingroup
\setlength{\unitlength}{0.0200bp}%
\begin{picture}(19800,11880)(0,0)%
\put(14541,5360){\makebox(0,0)[l]{\strut{}-0.69}}%
\put(14541,5965){\makebox(0,0)[l]{\strut{}-0.685}}%
\put(14541,6571){\makebox(0,0)[l]{\strut{}-0.68}}%
\put(14541,7177){\makebox(0,0)[l]{\strut{}-0.675}}%
\put(14541,7783){\makebox(0,0)[l]{\strut{}-0.67}}%
\put(14541,8389){\makebox(0,0)[l]{\strut{}-0.665}}%
\put(14541,8995){\makebox(0,0)[l]{\strut{}-0.66}}%
\put(9900,1122){\makebox(0,0){\strut{}$p$ [kbar]}}%
\put(4782,6190){\rotatebox{90}{\makebox(0,0){\strut{}$c$}}}%
\put(6966,1872){\makebox(0,0){\strut{} 0}}%
\put(8433,1872){\makebox(0,0){\strut{} 20}}%
\put(9900,1872){\makebox(0,0){\strut{} 40}}%
\put(11367,1872){\makebox(0,0){\strut{} 60}}%
\put(12834,1872){\makebox(0,0){\strut{} 80}}%
\put(5907,2522){\makebox(0,0)[r]{\strut{} 0}}%
\put(5907,3989){\makebox(0,0)[r]{\strut{} 0.2}}%
\put(5907,5457){\makebox(0,0)[r]{\strut{} 0.4}}%
\put(5907,6923){\makebox(0,0)[r]{\strut{} 0.6}}%
\put(5907,8391){\makebox(0,0)[r]{\strut{} 0.8}}%
\put(5907,9858){\makebox(0,0)[r]{\strut{} 1}}%
\end{picture}%
\endgroup
\\[-0.6cm]
\hspace*{6cm}$T$=800K\\[-1.2cm]
\begin{picture}(0,0)%
\includegraphics{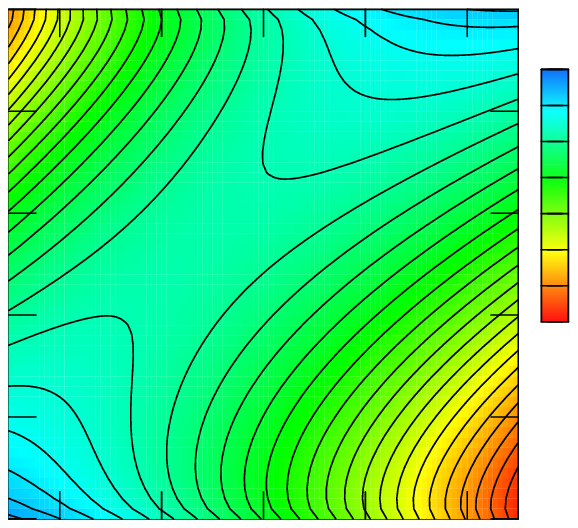}%
\end{picture}%
\begingroup
\setlength{\unitlength}{0.0200bp}%
\begin{picture}(19800,11880)(0,0)%
\put(14541,5360){\makebox(0,0)[l]{\strut{}-0.688}}%
\put(14541,5879){\makebox(0,0)[l]{\strut{}-0.686}}%
\put(14541,6398){\makebox(0,0)[l]{\strut{}-0.684}}%
\put(14541,6917){\makebox(0,0)[l]{\strut{}-0.682}}%
\put(14541,7437){\makebox(0,0)[l]{\strut{}-0.68}}%
\put(14541,7956){\makebox(0,0)[l]{\strut{}-0.678}}%
\put(14541,8475){\makebox(0,0)[l]{\strut{}-0.676}}%
\put(14541,8995){\makebox(0,0)[l]{\strut{}-0.674}}%
\put(9900,1122){\makebox(0,0){\strut{}$p$ [kbar]}}%
\put(4782,6190){\rotatebox{90}{\makebox(0,0){\strut{}$c$}}}%
\put(6966,1872){\makebox(0,0){\strut{} 0}}%
\put(8433,1872){\makebox(0,0){\strut{} 20}}%
\put(9900,1872){\makebox(0,0){\strut{} 40}}%
\put(11367,1872){\makebox(0,0){\strut{} 60}}%
\put(12834,1872){\makebox(0,0){\strut{} 80}}%
\put(5907,2522){\makebox(0,0)[r]{\strut{} 0}}%
\put(5907,3989){\makebox(0,0)[r]{\strut{} 0.2}}%
\put(5907,5457){\makebox(0,0)[r]{\strut{} 0.4}}%
\put(5907,6923){\makebox(0,0)[r]{\strut{} 0.6}}%
\put(5907,8391){\makebox(0,0)[r]{\strut{} 0.8}}%
\put(5907,9858){\makebox(0,0)[r]{\strut{} 1}}%
\end{picture}%
\endgroup
\\[-0.6cm]
\hspace*{6cm}$T$=1600K\\[-1.2cm]
\begin{picture}(0,0)%
\includegraphics{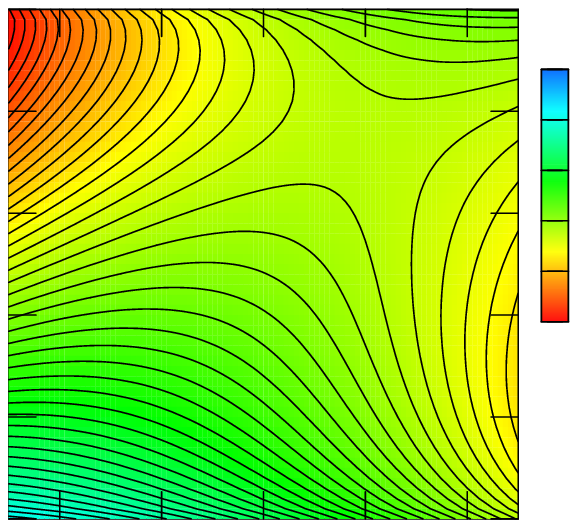}%
\end{picture}%
\begingroup
\setlength{\unitlength}{0.0200bp}%
\begin{picture}(19800,11880)(0,0)%
\put(14541,5360){\makebox(0,0)[l]{\strut{}-0.7}}%
\put(14541,6087){\makebox(0,0)[l]{\strut{}-0.695}}%
\put(14541,6814){\makebox(0,0)[l]{\strut{}-0.69}}%
\put(14541,7541){\makebox(0,0)[l]{\strut{}-0.685}}%
\put(14541,8268){\makebox(0,0)[l]{\strut{}-0.68}}%
\put(14541,8995){\makebox(0,0)[l]{\strut{}-0.675}}%
\put(9900,1122){\makebox(0,0){\strut{}$p$ [kbar]}}%
\put(4782,6190){\rotatebox{90}{\makebox(0,0){\strut{}$c$}}}%
\put(6966,1872){\makebox(0,0){\strut{} 0}}%
\put(8433,1872){\makebox(0,0){\strut{} 20}}%
\put(9900,1872){\makebox(0,0){\strut{} 40}}%
\put(11367,1872){\makebox(0,0){\strut{} 60}}%
\put(12834,1872){\makebox(0,0){\strut{} 80}}%
\put(5907,2522){\makebox(0,0)[r]{\strut{} 0}}%
\put(5907,3989){\makebox(0,0)[r]{\strut{} 0.2}}%
\put(5907,5457){\makebox(0,0)[r]{\strut{} 0.4}}%
\put(5907,6923){\makebox(0,0)[r]{\strut{} 0.6}}%
\put(5907,8391){\makebox(0,0)[r]{\strut{} 0.8}}%
\put(5907,9858){\makebox(0,0)[r]{\strut{} 1}}%
\end{picture}%
\endgroup
\\[-1.2cm]
\end{minipage}
\caption{(color) Gibbs free energies for $T$=0, 800 and 1600K. 
In order to enhance the readability of the plots
the energies have been calibrated by a linear term, proportional to pressure.
\label{Gibbs:figs}
}
\end{figure}

Although the concentration-dependent free energies are the quantities
directly accessible from the calculations, they do not easily reveal
the full information on the phase diagram.  In order to determine the
full $p-T$ phase diagram, it is necessary to calculate the Gibbs free
energy:
\begin{equation}
\label{eq:Gibbs}
G(T,c,p) = F(T,c,V(T,c,p)) + p V(T,c,p) .
\end{equation}
The Gibbs free energies are displayed in Fig. \ref{Gibbs:figs}.
From them, at each given pressure and temperature, we can determine
the concentration of the trivalent Ce, by minimizing the Gibbs free
energy with respect to $c$. At zero temperature one finds (for low
pressures) two local minima, associated with $c=0$ and $c=1$. By increasing the
pressure, the order of the minima changes and the minimizing
concentration jumps from $1$ to $0$. At higher temperatures the minima
start moving towards intermediate concentrations. Only above the
critical temperature, one finds the minimum smoothly changing from 
low to high concentrations.

\begin{figure} 
\begin{picture}(0,0)%
\includegraphics{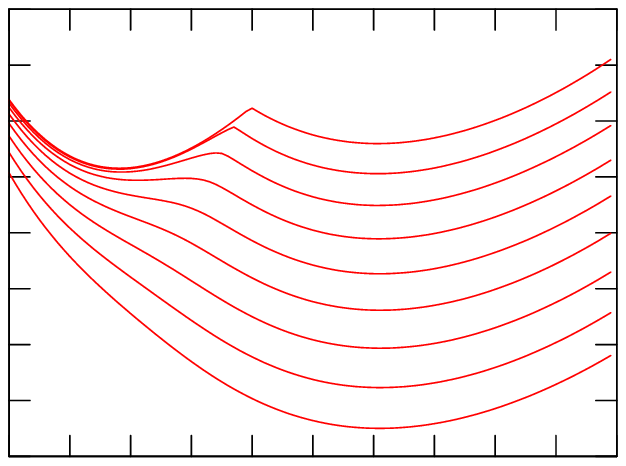}%
\end{picture}%
\begingroup
\setlength{\unitlength}{0.0200bp}%
\begin{picture}(12599,8640)(0,0)%
\put(2750,1650){\makebox(0,0)[r]{\strut{}-0.7}}%
\put(2750,2455){\makebox(0,0)[r]{\strut{}-0.695}}%
\put(2750,3260){\makebox(0,0)[r]{\strut{}-0.69}}%
\put(2750,4065){\makebox(0,0)[r]{\strut{}-0.685}}%
\put(2750,4870){\makebox(0,0)[r]{\strut{}-0.68}}%
\put(2750,5675){\makebox(0,0)[r]{\strut{}-0.675}}%
\put(2750,6480){\makebox(0,0)[r]{\strut{}-0.67}}%
\put(2750,7285){\makebox(0,0)[r]{\strut{}-0.665}}%
\put(2750,8090){\makebox(0,0)[r]{\strut{}-0.66}}%
\put(3025,1100){\makebox(0,0){\strut{} 140}}%
\put(3900,1100){\makebox(0,0){\strut{} 150}}%
\put(4775,1100){\makebox(0,0){\strut{} 160}}%
\put(5650,1100){\makebox(0,0){\strut{} 170}}%
\put(6525,1100){\makebox(0,0){\strut{} 180}}%
\put(7400,1100){\makebox(0,0){\strut{} 190}}%
\put(8275,1100){\makebox(0,0){\strut{} 200}}%
\put(9150,1100){\makebox(0,0){\strut{} 210}}%
\put(10025,1100){\makebox(0,0){\strut{} 220}}%
\put(10900,1100){\makebox(0,0){\strut{} 230}}%
\put(11775,1100){\makebox(0,0){\strut{} 240}}%
\put(550,4870){\rotatebox{90}{\makebox(0,0){\strut{}$F(T,V)$ [Ry]}}}%
\put(7400,275){\makebox(0,0){\strut{}$V$ [a.u.$^3$]}}%
\end{picture}%
\endgroup
\caption{\label{fig:Efree-eq} (color online)
The free energies as function of the volume for the temperatures 
0 (highest curve), 200, 400, 600, 800, 1000, 1200, 1400 and 1600 K (lowest curve).
A constant of 17717 Ry has been added to all energies.
}
\end{figure}

We can obtain the free energy of the physical system at a given volume 
by evaluating the concentration dependent free energy at the
minimizing  concentration $c_{\rm min}$:
\begin{equation}
F(T,V) = F(T,c_{\rm min},V) \,.
\end{equation}
These free energies are displayed in Fig.~\ref{fig:Efree-eq}, which clearly shows the
double-well behavior for low temperatures, which is gradually smoothened out with increasing
temperatures. Furthermore one finds that, at elevated temperatures, the free energy is mainly
lowered at large lattice constants, corresponding to the $\gamma$-phase, with its larger entropy.

\begin{figure}[b]
\begin{picture}(10,5)
\put(0.1,2.9){\rotatebox[origin=center]{90}{$p$ [kbar]}}
\put(3.8,0){$V$ [a.u.$^{3}$]}
\put(0.5,0.15){\includegraphics[scale=0.75]{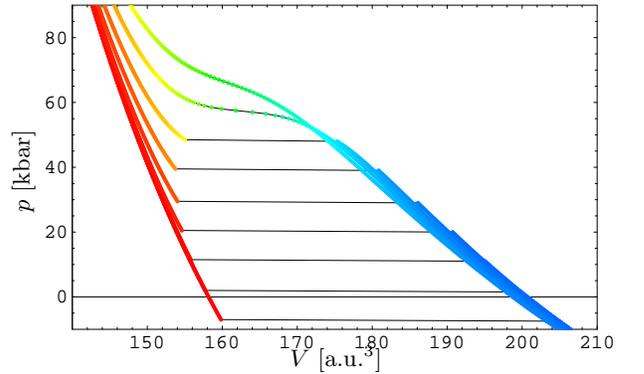}}
\end{picture}
\caption{\label{Ce-p-V} (color)
  Calculated isotherms for the temperatures
  T=0 (lowest curve), 200, 400, 600, 800, 1000, 1200, 1400, and, 1600K (highest curve).  The color
  indicates the fraction of localized electrons: blue is all localized
  ($\gamma$-phase) and red is all delocalized ($\alpha$-phase).} 
\end{figure}
Inserting the minimizing concentration $c_{\rm min}$ into the pressure-volume
relation
\begin{equation}
\label{eq:p-V}
p(T,V) = p(T,c_{\rm min},V) = -\frac{\partial}{\partial V} F(T,c_{\rm min},V) 
\end{equation}
allows to calculate the isotherms of Ce, which are displayed in
Fig. \ref{Ce-p-V}. It can be seen that the average valence, close to the
coexistence line, gradually changes with increasing temperature. Above the
critical temperature, the valence changes continuously with increasing pressure
from trivalent to tetravalent. 

\begin{figure}[h] 
\hspace{-7.0cm}
\begin{minipage}{7cm}
\vspace*{-1.2cm}
\begin{picture}(0,0)%
\includegraphics{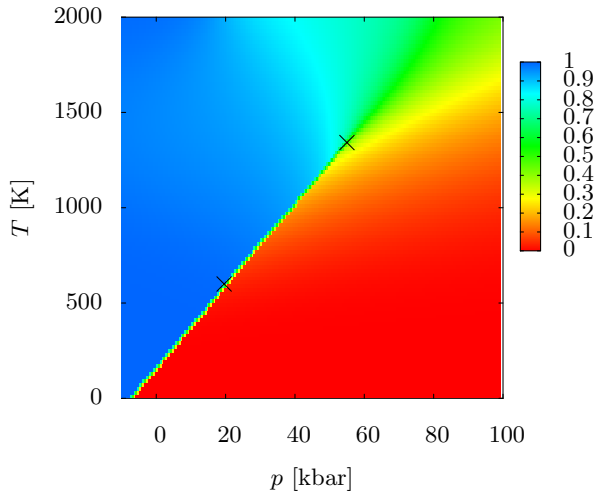}%
\end{picture}%
\begingroup
\setlength{\unitlength}{0.0200bp}%
\begin{picture}(19800,11880)(0,0)%
\put(14477,5402){\makebox(0,0)[l]{\strut{} 0}}%
\put(14477,5758){\makebox(0,0)[l]{\strut{} 0.1}}%
\put(14477,6114){\makebox(0,0)[l]{\strut{} 0.2}}%
\put(14477,6470){\makebox(0,0)[l]{\strut{} 0.3}}%
\put(14477,6826){\makebox(0,0)[l]{\strut{} 0.4}}%
\put(14477,7182){\makebox(0,0)[l]{\strut{} 0.5}}%
\put(14477,7538){\makebox(0,0)[l]{\strut{} 0.6}}%
\put(14477,7894){\makebox(0,0)[l]{\strut{} 0.7}}%
\put(14477,8250){\makebox(0,0)[l]{\strut{} 0.8}}%
\put(14477,8606){\makebox(0,0)[l]{\strut{} 0.9}}%
\put(14477,8962){\makebox(0,0)[l]{\strut{} 1}}%
\put(9900,1081){\makebox(0,0){\strut{}$p$ [kbar]}}%
\put(4437,6215){\rotatebox{90}{\makebox(0,0){\strut{}$T$ [K]}}}%
\put(6960,1906){\makebox(0,0){\strut{} 0}}%
\put(8267,1906){\makebox(0,0){\strut{} 20}}%
\put(9574,1906){\makebox(0,0){\strut{} 40}}%
\put(10880,1906){\makebox(0,0){\strut{} 60}}%
\put(12187,1906){\makebox(0,0){\strut{} 80}}%
\put(13494,1906){\makebox(0,0){\strut{} 100}}%
\put(5949,2621){\makebox(0,0)[r]{\strut{} 0}}%
\put(5949,4418){\makebox(0,0)[r]{\strut{} 500}}%
\put(5949,6215){\makebox(0,0)[r]{\strut{} 1000}}%
\put(5949,8012){\makebox(0,0)[r]{\strut{} 1500}}%
\put(5949,9809){\makebox(0,0)[r]{\strut{} 2000}}%
\end{picture}%
\endgroup
\end{minipage}
\caption{(color) Phase diagram obtained for the pseudoalloy, composed of $\alpha$-
  and $\gamma$-Ce.  The crosses indicate the calculated and 
  experimental critical points. The color
  indicates the fraction of localized electrons: blue is all localized
  ($\gamma$-phase) and red is all delocalized ($\alpha$-phase).
\label{full-phase-diag}}
\end{figure}

In Fig. \ref{full-phase-diag} we present the phase diagram, obtained
from the free energies of the $\alpha$-$\gamma$ pseudoalloy, with 
the $\gamma$-phase described by the DLM approach. It can clearly be seen
in the figure how the transition becomes continuous above the
critical temperature. The experimentally observed critical point (600K, 20 kbar)
falls on top of the calculated phase separation line, which starts at the zero
temperature transition pressure of -7.4 kbar. This means that the slope of the
phase separation line is in very good agreement with experiments. The critical
temperature overestimates the experimental one by roughly a factor of 2,
which is still reasonable considering that the critical temperature 
is very sensitive to various small details of the calculations and in
particular the theoretical lattice parameters of both the Ce phases.
Note that the $T_c$ at zero pressure of 169 K (see Table \ref{crit-points}) compares 
well with the experimental value of 141$\pm$10 K.

\begin{figure}
\begin{picture}(0,0)%
\includegraphics{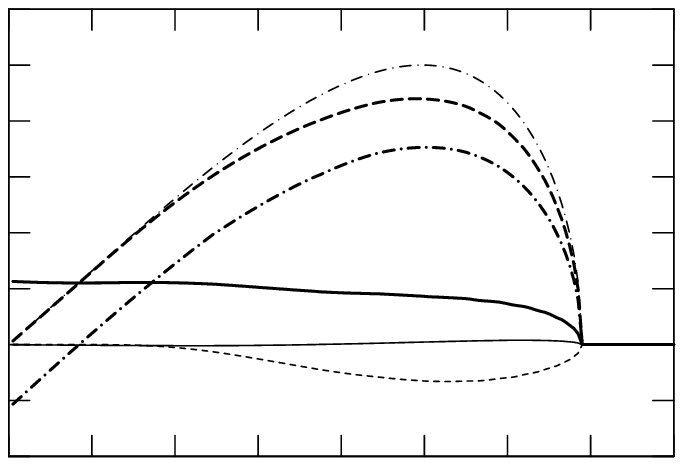}%
\end{picture}%
\begingroup
\setlength{\unitlength}{0.0200bp}%
\begin{picture}(12599,8640)(0,0)%
\put(1925,1650){\makebox(0,0)[r]{\strut{}-4}}%
\put(1925,2455){\makebox(0,0)[r]{\strut{}-2}}%
\put(1925,3260){\makebox(0,0)[r]{\strut{} 0}}%
\put(1925,4065){\makebox(0,0)[r]{\strut{} 2}}%
\put(1925,4870){\makebox(0,0)[r]{\strut{} 4}}%
\put(1925,5675){\makebox(0,0)[r]{\strut{} 6}}%
\put(1925,6480){\makebox(0,0)[r]{\strut{} 8}}%
\put(1925,7285){\makebox(0,0)[r]{\strut{} 10}}%
\put(1925,8090){\makebox(0,0)[r]{\strut{} 12}}%
\put(2200,1100){\makebox(0,0){\strut{} 0}}%
\put(3397,1100){\makebox(0,0){\strut{} 200}}%
\put(4594,1100){\makebox(0,0){\strut{} 400}}%
\put(5791,1100){\makebox(0,0){\strut{} 600}}%
\put(6988,1100){\makebox(0,0){\strut{} 800}}%
\put(8184,1100){\makebox(0,0){\strut{} 1000}}%
\put(9381,1100){\makebox(0,0){\strut{} 1200}}%
\put(10578,1100){\makebox(0,0){\strut{} 1400}}%
\put(11775,1100){\makebox(0,0){\strut{} 1600}}%
\put(550,4870){\rotatebox{90}{\makebox(0,0){\strut{}$\Delta E$ [mRy]}}}%
\put(6987,275){\makebox(0,0){\strut{}$T$ [K]}}%
\end{picture}%
\endgroup
\caption{
\label{fig:jumps}
Discontinuities of the total energy (thick solid line), the total
entropy $T S$ (thick dashed line) and the $p V$ term (thick
dashed-dotted line) over the phase separation line as function of the
temperature. The entropy term is further decomposed into the
electronic (thin solid line), the mixing (thin dashed line) and the
magnetic (thin dashed-dotted line) contribution.}
\end{figure}

Finally we examine in more detail the discontinuity across the phase
separation line. Figure \ref{fig:jumps} shows the magnitude of the
discontinuities for the various ingredients of the Gibbs free energy.
As expected, all contributions vanish at the critical temperature,
above which there is a continuous cross-over between the $\alpha$- and
the $\gamma$-phase.  It also can be seen from this figure that the
entropy discontinuity is by far the largest contribution. The phase
transition is therefore driven by entropy, rather than by energetics.
The entropy discontinuity itself is mainly determined by the magnetic
entropy.


\subsection{Analysis of results}
\label{subsec:analysis}

In order to investigate the importance of the different aspects of
the present calculations (i.e. the DLM description of the $\gamma$-phase, the
inclusion of finite temperature effects in the electronic free energy,
and the CPA itself), as compared to earlier studies,
we have also performed a set of calculations where, 
selectively, we neglect some of these effects and look at the consequences. 
In particular, we study the influence of these effects on the critical
temperature and the slope of the phase separation line.
The results of these calculations are summarized in Table
\ref{crit-points}, in comparison with the results of earlier theoretical,
as well as, experimental studies.

\begin{table*}
\caption{
\label{crit-points}
The critical temperature and pressure, as well as, the zero temperature  and room temperature 
transition pressures and the zero pressure transition temperature  for different 
calculations: DLM and Ferro refer to the disordered or ferromagnetic alignment of the local 
moments in the $\gamma$-phase, CF and SO indicate the crystal field or 
spin-orbit scenario, as discussed in the text. The index
I denotes calculations with finite temperature effects included in the band-structure and
the CPA, II refers to the neglect of these finite temperature effects, and III represents
calculations, where in addition the concentration dependence was approximated by
a linear interpolation. The main results, which are also shown in the figures, are the
DLM-CF (I) calculations, printed as bold in the table.}
\begin{tabular}{l|ccccccccccccccccc}
\hline \hline
 & \multicolumn{3}{c}{DLM-CF} & \multicolumn{3}{c}{DLM-SO} &
\multicolumn{3}{c}{Ferro-CF} & \multicolumn{3}{c}{Ferro-SO} &
Svane$^a$ & Johansson$^b$ & KVC$^c$ & Prom$^d$ & exp. \\
 & I & II & III  & I & II & III  & I & II & III  & I & II & III \\
\hline
$T_c$       & {\bf 1377} & 1528 & 1129 & 1407 & 1568 & 1157 & 1444 & 1660 & 1139 & 1471 & 1689 & 1166 & 1300 & 980 & 520 & 600 & 600 \\
$p(T_c)$    & {\bf 56}   &   62 &   51 &   47 &   52 &   43 &   64 &   74 &   58 &   53 &   61 &   49 &   47 &  39 &  39 &  18 &  20 \\
$p(T=0K)$   & {\bf -7.4} & -7.4 & -7.4 & -7.4 & -7.4 & -7.4 & -2.3 & -2.3 & -2.3 & -2.2 & -2.3 & -2.3 & -1.0 &  -6 &  -6 &     &  -7 \\
$p(T=300K)$ & {\bf 6.1}  &  6.2 &  6.2 &  4.1 &  4.2 &  4.2 & 11.0 & 11.4 & 11.4 &  9.0 &  9.5 &  9.5 &   10 &   7 &   8 &   6 &   6 \\
$T_c(p=0)$  & {\bf 169}  &  167 &  167 &  196 &  194 &  194 &   52 &   52 &   52 &   61 &   61 &   61 &      & 135 &     &     & 140$\pm$10 \\
\hline \hline
\end{tabular}
\begin{flushleft}
$^a$ Ref. \onlinecite{Svane:96} \\
$^b$ Ref. \onlinecite{JohanssonEtAl:95}\\
$^c$ Kondo volume collapse model, Ref. \onlinecite{AllenLiu:92}\\
$^d$ Promotion model, Ref. \onlinecite{CoqblinBlandin:68}
\end{flushleft}
\end{table*}

First we focus on the importance of the disordered local moments representation 
of the $\gamma$-phase.  By comparing the DLM-results in Table~\ref{crit-points} 
with those marked by 'Ferro' (ferromagnetic
calculations for the $\gamma$-phase), one finds that the DLM calculations 
lead to a moderate
lowering of the critical temperature, and a more negative zero
temperature transition pressure.  This can easily be understood,
since at zero temperature, the ferromagnetically ordered phase has a
lower energy as compared to the disordered phase. Experimentally such
a magnetic order is not observed, since at low temperatures (and
positive pressures) Ce is in its non-magnetic $\alpha$-phase.  The
lowering of the critical temperature cannot be easily identified with
a specific aspect of the DLM calculations, since many effects, such as
the curvature of the free energies with respect to the concentration,
but also the anharmonic terms in the total energy as a function of
volume, conspire to determine the phase diagram. 

\begin{figure}
\begin{picture}(0,0)%
\includegraphics{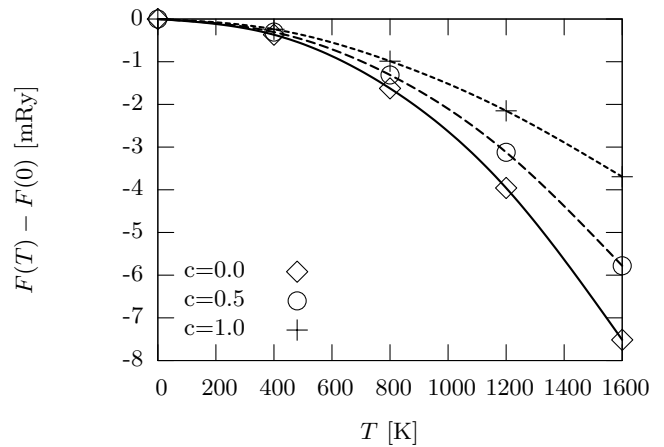}%
\end{picture}%
\begingroup
\setlength{\unitlength}{0.0200bp}%
\begin{picture}(12599,8640)(0,0)%
\put(2750,1650){\makebox(0,0)[r]{\strut{}-8}}%
\put(2750,2455){\makebox(0,0)[r]{\strut{}-7}}%
\put(2750,3260){\makebox(0,0)[r]{\strut{}-6}}%
\put(2750,4065){\makebox(0,0)[r]{\strut{}-5}}%
\put(2750,4870){\makebox(0,0)[r]{\strut{}-4}}%
\put(2750,5675){\makebox(0,0)[r]{\strut{}-3}}%
\put(2750,6480){\makebox(0,0)[r]{\strut{}-2}}%
\put(2750,7285){\makebox(0,0)[r]{\strut{}-1}}%
\put(2750,8090){\makebox(0,0)[r]{\strut{} 0}}%
\put(3025,1100){\makebox(0,0){\strut{} 0}}%
\put(4119,1100){\makebox(0,0){\strut{} 200}}%
\put(5213,1100){\makebox(0,0){\strut{} 400}}%
\put(6306,1100){\makebox(0,0){\strut{} 600}}%
\put(7400,1100){\makebox(0,0){\strut{} 800}}%
\put(8494,1100){\makebox(0,0){\strut{} 1000}}%
\put(9588,1100){\makebox(0,0){\strut{} 1200}}%
\put(10681,1100){\makebox(0,0){\strut{} 1400}}%
\put(11775,1100){\makebox(0,0){\strut{} 1600}}%
\put(550,4870){\rotatebox{90}{\makebox(0,0){\strut{}$F(T)-F(0)$ [mRy]}}}%
\put(7400,275){\makebox(0,0){\strut{}$T$ [K]}}%
\put(4675,3325){\makebox(0,0)[r]{\strut{}c=0.0}}%
\put(4675,2775){\makebox(0,0)[r]{\strut{}c=0.5}}%
\put(4675,2225){\makebox(0,0)[r]{\strut{}c=1.0}}%
\end{picture}%
\endgroup
\caption{
\label{finite-T-diff}
Temperature dependence of the electronic free energy as defined in the text.}
\end{figure}

The second point is the effect of finite temperature on the total energy
and the electronic entropy. In earlier studies of the phase diagram,
\cite{Svane:94,Svane:96,JohanssonEtAl:95} the electronic structure
calculations were performed at zero temperature, and the 
finite temperature effect entered only via the mixing entropy, the magnetic
entropy and in Ref.~\onlinecite{JohanssonEtAl:95} also
the vibrational entropy. This means that $E_{\rm tot}(T,c,V)$ was replaced by $E_{\rm tot}(0,c,V)$ 
and the electronic entropy $S_{\rm el}$ was neglected.
In Fig. \ref{finite-T-diff}, we analyze the difference between the
electronic free energy $F_{\rm el}(T,c,V) = E_{\rm tot}(T,c,V) - T S_{\rm
  el}(T,c,V)$ and the total energy at $T=0$.
The difference $F_{\rm el}(T,c,V) - F_{\rm el}(0,c,V)$ exhibits a
moderate dependence on the concentration. The larger effect for the
$\alpha$-phase is easily explained using a low temperature expansion.
The main effect of the finite temperatures is the broadening of the
Fermi function. To lowest order in temperature, the change of the free
energy is proportional to the density of states at the Fermi level.
The effect on the phase diagram can be seen in Table \ref{crit-points} by
comparing the columns I and II. Neglecting these finite temperature
effects gives rise to an increase of the critical temperature by roughly
200K, while the slope of the phase separation line remains unaltered.

Next is the effect of the CPA. In the study by Svane,
\cite{Svane:96} it was suggested that the weak departure from linearity
of the total energy curves as a function of concentration (Fig.
\ref{CPA-fig}) did not play an important role. Thus, we recalculated
the phase diagram, replacing the full concentration dependence of the
total energy by the linear interpolation
\begin{equation}
\label{eq:2}
E_{\rm lin}(c,V) = (1-c) E(0,V) + c E(1,V).
\end{equation}
The effect of this approximation can be seen in Table
\ref{crit-points} by comparing the columns II and III, respectively.
Both sets of calculations use the $T=0$K total energies only.  The
transition temperature obtained from the linear interpolation is
strikingly reduced, in comparison to the CPA calculation, and similar
to the one reported by Svane.  As pointed out by Johansson {\em et
  al.},\cite{JohanssonEtAl:95} the critical temperature is to a large
extent determined by the mixing entropy.  Without the mixing entropy
one would at all temperatures find a sudden transition (with a finite
volume collapse) between the low and high pressure
phases.
The mixing entropy will, if the temperature is high enough, lead to a
minimum of the free energy for an intermediate concentration of the
$\gamma$-phase, eventually resulting in a continuous crossover
between the low-pressure and the high pressure phase.
As seen in Fig. \ref{CPA-fig}, the CPA gives rise to a convex (from
above) curvature of the total energies. It effectively reduces the
mixing entropy and increases the critical temperature.
It can also be seen that within this linear approximation, the difference
between the DLM and the 'Ferro' calculations is strongly reduced, which means
that the main effect of DLM is not only the energy lowering of the $\gamma$-phase,
but more importantly a different shape of the energy versus concentration
curves.

Last, we discuss the effect of the degeneracy of the
$\gamma$-phase ground state. As described above, in the CF
calculations a degeneracy of 8 was used, corresponding to almost degenerate
$A_{2u}$ and $T_{2u}$ states, arising from a crystal field
splitting. The results to be compared here are shown in the
columns CF and SO of Table \ref{crit-points}. In the SO
scenario, the term proportional to $\ln 6$, corresponding to a $J=5/2$ state, 
was used for the magnetic entropy, instead of $\ln 8$, as in the CF
scenario.  One can see that the magnetic entropy, being linear in
$c$, determines mostly the slope of the phase separation line and has
only a minor effect on the magnitude of the critical temperature.  The
results obtained with the DLM and the magnetic entropy due to the CF
splitting show better agreement of the calculated transition
pressures with experiment. This, however, may be due to a cancellation of
errors. The slope of the phase separation line is mainly given by
the ratio $(S_{\gamma}-S_{\alpha})/(V_{\gamma}-V_{\alpha})$, where
$S_{\alpha(\gamma)}$ and $V_{\alpha(\gamma)}$ are the the T=0 values of the
entropy and the volume of the $\alpha(\gamma)$-phase.

Due to the
larger underestimation of the volume of the $\alpha$-phase, the volume collapse
is slightly overestimated in our approach. Therefore the higher value of the
entropy difference in the CF scenario, as opposed to the SO scenario, 
leads to a better agreement with the experimental slope.
It should be noted though that these
calculations do not include the vibrational entropy.
To  estimate the effect of this vibrational entropy, we
recalculate the phase diagram using a simple model for
the vibrational entropy, namely
\begin{equation}
\label{eq:3}
S_{\rm vib}(c) = -k_B \, c \, \Delta S_{\rm vib}^{\gamma-\alpha} ,
\end{equation}
with the value of $\Delta S_{\rm vib}^{\gamma-\alpha} \approx 0.75$,
as suggested by Jeong {\em et al.}\cite{JeongEtAl:04} In doing so, the
critical temperature is only slightly reduced to 1292 K, while the
critical pressure increases to 69 kbar. In view of the above
discussion, it is not surprising that a contribution, which is purely
linear in $c$, basically affects only the critical pressure.  We
conclude from this result that this simplified model for the
vibrational entropy is too crude and a more sophisticated one should
be put in place. One should here add that Johansson {\em et
al.}\cite{JohanssonEtAl:95} have employed a Debye-Gr{\"u}neisen
model, while Svane\cite{Svane:96} has completely neglected the vibrational entropy.

Finally, we would like to comment on the finding of
Niklasson {\em et al.}\cite{Niklasson:03} that 
the disordered local moments can give rise to a localization of f-electrons.
From our DLM only calculation, without applying SIC, we found that it is
not possible to stabilize a local moment in Ce, except at very large volumes,
where even the LSD yields a magnetic solution.


\section{Discussion}
\label{sec:discussion}
As already mentioned, experiments indicate that the $\alpha$-phase is
not composed of tetravalent Ce atoms, but is rather described by an intermediate valence
of 3.67.\cite{KoskenmakiGschneidner:78}
From Fig. \ref{free-energy-plots}, we see that an intermediate
valence, i.e., a fractional concentration ($0<c<1$),  would lead to
an increased equilibrium lattice constant. 
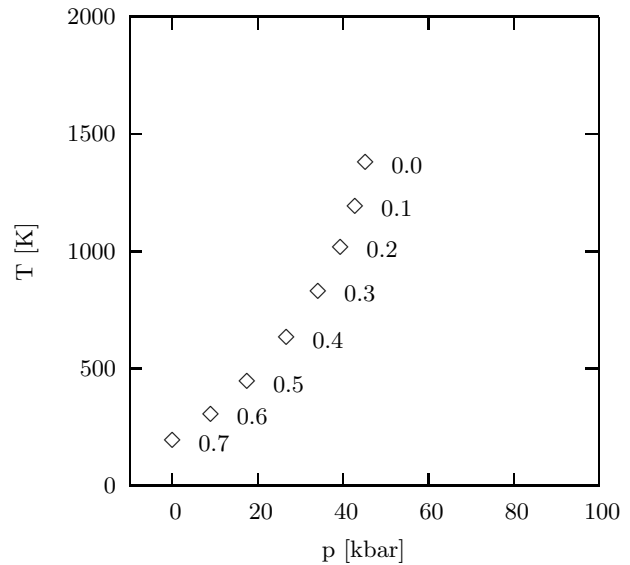
\begin{figure}
\hspace{-2.0cm}
\begin{minipage}{7cm}
\setlength{\unitlength}{0.240900pt}
\ifx\plotpoint\undefined\newsavebox{\plotpoint}\fi
\sbox{\plotpoint}{\rule[-0.200pt]{0.400pt}{0.400pt}}%
\begin{picture}(1500,900)(0,0)
\sbox{\plotpoint}{\rule[-0.200pt]{0.400pt}{0.400pt}}%
\put(201.0,123.0){\rule[-0.200pt]{4.818pt}{0.400pt}}
\put(181,123){\makebox(0,0)[r]{ 0}}
\put(918.0,123.0){\rule[-0.200pt]{4.818pt}{0.400pt}}
\put(201.0,307.0){\rule[-0.200pt]{4.818pt}{0.400pt}}
\put(181,307){\makebox(0,0)[r]{ 500}}
\put(918.0,307.0){\rule[-0.200pt]{4.818pt}{0.400pt}}
\put(201.0,491.0){\rule[-0.200pt]{4.818pt}{0.400pt}}
\put(181,491){\makebox(0,0)[r]{ 1000}}
\put(918.0,491.0){\rule[-0.200pt]{4.818pt}{0.400pt}}
\put(201.0,676.0){\rule[-0.200pt]{4.818pt}{0.400pt}}
\put(181,676){\makebox(0,0)[r]{ 1500}}
\put(918.0,676.0){\rule[-0.200pt]{4.818pt}{0.400pt}}
\put(201.0,860.0){\rule[-0.200pt]{4.818pt}{0.400pt}}
\put(181,860){\makebox(0,0)[r]{ 2000}}
\put(918.0,860.0){\rule[-0.200pt]{4.818pt}{0.400pt}}
\put(268.0,123.0){\rule[-0.200pt]{0.400pt}{4.818pt}}
\put(268,82){\makebox(0,0){ 0}}
\put(268.0,840.0){\rule[-0.200pt]{0.400pt}{4.818pt}}
\put(402.0,123.0){\rule[-0.200pt]{0.400pt}{4.818pt}}
\put(402,82){\makebox(0,0){ 20}}
\put(402.0,840.0){\rule[-0.200pt]{0.400pt}{4.818pt}}
\put(536.0,123.0){\rule[-0.200pt]{0.400pt}{4.818pt}}
\put(536,82){\makebox(0,0){ 40}}
\put(536.0,840.0){\rule[-0.200pt]{0.400pt}{4.818pt}}
\put(670.0,123.0){\rule[-0.200pt]{0.400pt}{4.818pt}}
\put(670,82){\makebox(0,0){ 60}}
\put(670.0,840.0){\rule[-0.200pt]{0.400pt}{4.818pt}}
\put(804.0,123.0){\rule[-0.200pt]{0.400pt}{4.818pt}}
\put(804,82){\makebox(0,0){ 80}}
\put(804.0,840.0){\rule[-0.200pt]{0.400pt}{4.818pt}}
\put(938.0,123.0){\rule[-0.200pt]{0.400pt}{4.818pt}}
\put(938,82){\makebox(0,0){ 100}}
\put(938.0,840.0){\rule[-0.200pt]{0.400pt}{4.818pt}}
\put(201.0,123.0){\rule[-0.200pt]{177.543pt}{0.400pt}}
\put(938.0,123.0){\rule[-0.200pt]{0.400pt}{177.543pt}}
\put(201.0,860.0){\rule[-0.200pt]{177.543pt}{0.400pt}}
\put(201.0,123.0){\rule[-0.200pt]{0.400pt}{177.543pt}}
\put(40,491){\rotatebox{90}{\makebox(0,0){T [K]}}}
\put(569,21){\makebox(0,0){p [kbar]}}
\put(572,630){\raisebox{-.8pt}{\makebox(0,0){$\Diamond$}}}
\put(572,630){\raisebox{-.8pt}{\makebox(0,0){\hspace*{1cm} 0.0}}}
\put(556,562){\raisebox{-.8pt}{\makebox(0,0){$\Diamond$}}}
\put(556,562){\raisebox{-.8pt}{\makebox(0,0){\hspace*{1cm} 0.1}}}
\put(533,497){\raisebox{-.8pt}{\makebox(0,0){$\Diamond$}}}
\put(533,497){\raisebox{-.8pt}{\makebox(0,0){\hspace*{1cm} 0.2}}}
\put(498,428){\raisebox{-.8pt}{\makebox(0,0){$\Diamond$}}}
\put(498,428){\raisebox{-.8pt}{\makebox(0,0){\hspace*{1cm} 0.3}}}
\put(448,355){\raisebox{-.8pt}{\makebox(0,0){$\Diamond$}}}
\put(448,355){\raisebox{-.8pt}{\makebox(0,0){\hspace*{1cm} 0.4}}}
\put(386,286){\raisebox{-.8pt}{\makebox(0,0){$\Diamond$}}}
\put(386,286){\raisebox{-.8pt}{\makebox(0,0){\hspace*{1cm} 0.5}}}
\put(329,235){\raisebox{-.8pt}{\makebox(0,0){$\Diamond$}}}
\put(329,235){\raisebox{-.8pt}{\makebox(0,0){\hspace*{1cm} 0.6}}}
\put(269,193){\raisebox{-.8pt}{\makebox(0,0){$\Diamond$}}}
\put(269,193){\raisebox{-.8pt}{\makebox(0,0){\hspace*{1cm} 0.7}}}
\put(201.0,123.0){\rule[-0.200pt]{177.543pt}{0.400pt}}
\put(938.0,123.0){\rule[-0.200pt]{0.400pt}{177.543pt}}
\put(201.0,860.0){\rule[-0.200pt]{177.543pt}{0.400pt}}
\put(201.0,123.0){\rule[-0.200pt]{0.400pt}{177.543pt}}
\end{picture}

\end{minipage}
\caption{Critical points of the phase transition, obtained when the concentration
of trivalent Ce atoms in the $\alpha$-phase is artificially fixed
at a finite $c_0$. The points are marked by their corresponding
value of $c_0$, where $c_0=0$ represents purely tetravalent Ce.
\label{int-valence}}
\end{figure}
One can simulate the effect of the intermediate valence for the
$\alpha$-phase by simply rescaling the concentrations when evaluating
the phase diagram. In Fig. \ref{int-valence} we present the
critical points of the phase transition obtained when we represent the
$\alpha$-phase by a non-zero concentration, $c_0$, of the trivalent Ce
atoms, added into the host of tetravalent Ce atoms. 
As can be seen in the figure, the critical temperature quickly
decreases with the increase of the admixture of trivalent Ce
in the $\alpha$-phase. In this calculation, the SIC-LSD energies have
been uniformly calibrated to keep the zero temperature transition
pressure at its original value of -7.4 kbar.  It can be seen in the
figure that the best value for critical temperature is obtained for
$c_0 \approx 0.4$, which corresponds to an intermediate valence of
3.6.

The intermediate valence scenario for the $\alpha$-phase, as discussed above, 
could result from dynamical fluctuations. These fluctuations could
be realized by describing Ce as a two level system (TLS). We will elaborate
on this idea in section \ref{sec:outlook}.  A mechanism, based on the
dynamical interaction of two states will, quite generally, be more
effective if the two states are close in energy. Looking at the total
energy as a function of the lattice spacing and the concentration
(Fig.~\ref{free-energy-plots}), one finds that the pure $\alpha$- and
$\gamma$-solutions (LDA and SIC-LSD, respectively) are
degenerate close to $a = 8.9$ atomic units.  The interaction of these
two states will lower the total energy at these lattice constants, and
eventually might establish them as the global minimum.  The increased
lattice constant would be closer to experimental values.  Clearly such
a state would be neither described by full localization nor
delocalization of the $f$-electron, but would be better characterized
as an intermediate valence state. In the following section we will
outline how we envision this two-level system to work.

The next point we want to discuss is the effect of lattice
relaxations.  In the intermediate valence regime (as described by the
static CPA) there is a rather large size-mismatch between the
$\alpha$- and the $\gamma$-atoms, which may give rise to strong
internal strains.  Allowing for lattice relaxations, which are not
considered here, would give rise to an energy lowering for
intermediate concentrations and could lead to intermediate valence,
even in the static limit, and hence to the reduction of the critical
temperature.

Another factor that might have a significant influence on the phase 
diagram and its characteristics is associated with the single-site 
aspect of the CPA. Being a single-site theory, the CPA cannot deal
with 'order in disorder', namely with short range order (SRO) in 
the distribution of the 'alloy' configurations. However, even in the
disordered phase it must be important to distinguish situations
where the nearest neighbours are preferentially like atoms from
those where they are unlike atoms. Such short range order is expected
to lower the free energy of the disordered state and hence lower
T$_{c}$ of the $\alpha-\gamma$ phase transition. In addition, it should
influence the energetically favourable relative concentration of the
two different components of the alloy. Thus by taking SRO
into account one might be able to move beyond the primitive alloy
analogy and improve on the present LSIC-KKR-CPA approach. The way it 
could be accomplished is by implementing the nonlocal extention of the
KKR-CPA,\cite{JarrellEtAl:01, RowlandsEtAl:03, RowlandsEtAl:04} which would allow to
treat possible correlated valence fluctuations near the $\alpha-\gamma$ 
phase transition in Ce.

Finally, the phase diagram of Ce suggests that at negative pressure
there could exist a quantum critical point (QCP), i.e., a localization
- delocalization transition at zero temperature, driven by pressure.
The vicinity of this QCP, although it is not accessible
experimentally, could still influence the physics of the material in
the accessible positive pressure range. The quantum fluctuations,
which are responsible for the transition, should also be visible in
its vicinity and could explain the correlated nature of the $\alpha$-
phase.

\section{Outlook}
\label{sec:outlook}
Several times in this article we have referred to the possibility of
going beyond the theory outlined above by allowing for dynamical valence and
spin fluctuations.
In short, the suggestion is that, as above, we regard the
self-interaction corrected and the not so corrected version of the
local potential at a site, as corresponding to two states of the
'atom' and allow for such 'atom' to tunnel between the two states
$|a>$ and $|b>$, which would form a two-level system. Electrons
interacting with such TLS were already thoroughly studied in the
context of metallic glasses.\cite{BlackGyorffy:78} In the present
scenario such a procedure would address the valence fluctuations.  On
the other hand, identifying the possible spin states of the
SI-corrected system with the two levels \footnote{The idea of a
  two-level system could, of course, be generalized to N levels.}
would constitute a possible dynamical generalization of the DLM
formalism, taking into account dynamical spin fluctuations. This might
describe the Kondo screening of the local moments at low temperatures.

Many of the consequences of such interactions are by now well 
known.\cite{BookCoxZawadowski} In such studies the TLS is an atom tunneling
between two nearly degenerate positions in a metallic 
environment.\cite{BookCoxZawadowski} The physics is particularly interesting in
the case of 'assisted tunneling' where the TLS changes its state as an
electron scatters from it.
However, it should be stressed that in the cases where the atom
changes its position in the tunneling process, the TLS is external to
the electron system whilst in the proposed model it is designed to
capture the physics of a slowly changing collective degree of freedom
of the electrons themselves. In this sense our TLS is very much like
the DLM in para\-magnetic metals.

From the point of view of the present perspective the most directly
relevant work on external TLS's is that of Vladar {\em et al.}\cite{VladarEtAl:86}
  They show that, if the TLS couples two or more
angular momentum states, as the temperature is lowered the high
temperature 'pseudoalloy' phase gives way, at a characteristic
temperature T$_{K},$ to a ground state which is an orbital analogue of
Kondo singlets of magnetic impurities in metals. In the case of the
LDA/SIC-LSD TLS, this means, that at high temperatures, we would find the
pseudoalloy, as described above, while at low temperatures, the system
would be neither in the LDA, nor in the SIC-LSD state, but would be in an
intermediate valence ground state.
Clearly, such dynamical fluctuations may also help to reduce the
critical temperature if the generalized Kondo-temperature, associated
with the TLS, is not too small.

As a final remark concerning the dynamical generalization of our
disordered local valence calculations we note that the above model is
an analogue of Yuval and Anderson's 'Kondo Hamiltonian' approach to the
magnetic impurity problem, as opposed to the full dynamical
calculation based on the Anderson model, deployed for the same problem
by Hamann.\cite{Hamann:69,Hamann:70} The relevant point to stress is 
that these two
calculations yield, in the appropriately asymptotic, namely scaling,
regime the same results. Thus, they lend credit to the above proposed 
short-cut to a first principles DMFT treatment of our fluctuating
valence problem.

\section{conclusions}
\label{sec:conclusions}
We have presented a new, multiple scattering, implementation of the SIC-LSD 
formalism for solids within KKR band structure method, combined with the CPA 
description of intermediate valences. The method has been illustrated on 
the application to the Ce $\alpha-\gamma$ phase transition. The results
have been discussed in detail, highlighting the new functionality and potential 
of this approach owing to a better static description of spin and valence
fluctuations. The importance of all the different aspects of the formalism 
has been analysed in detail. This new method is not to be looked at as 
an alternative to the earlier implementations within the LMTO-ASA band
structure method. Its great potential, and in some way superiority, arises 
from the local and multiple scattering aspects through which the method
lends itself easily to various generalizations and extentions on the
account of the straightforward determination of the one-electron Green's
function. Of particular interest here is an inclusion of dynamical 
fluctuations, for which a roadmap has been briefly outlined in the previous section.
The results of the present paper constitute the first crucial steps on this
road towards dynamics.

Finally, there is one more aspect of the present results which warrants further comment. 
As we have reported in Sec.~\ref{subsec:analysis}, in our DLM calculations local moment 
formed only when the local electronic structure was described by SIC-LSD, whilst it
iterated to zero when LDA was used in recalculating the spin polarized crystal  
potential. This is precisely the behavior one would have expected on the basis 
of the numerous successes of SIC in predicting moment formation, and no moment 
formation in applications to extended systems.\cite{SzotekEtAl:94,Svane:96} Clearly, 
the fact that our local implementation of SIC, namely LSIC, behaves in this way 
lends strong support to our contention that self-interaction correction formalism
can be, and perhaps should be, applied at the local level.

\section{Acknowledgments}
This work was partially funded by the EU Research Training Network (contract: HPRN-CT-2002-00295)
'Ab-initio Computation of Electronic Properties of f-electron Materials' and by the Deutsche
Forschungsgemeinschaft through the Forschergruppe 'Oxidic Interfaces'.
Part of the calculations were performed at the Rechenzentrum Garching, as well as the
John von Neumann Institute for Computing in J{\"u}lich.
Ole Krog Andersen is gratefully acknowledged for many useful and enlightening descussion over 
the years, and in particular those on a possible use of NMTOs in the full SIC-LSD 
implementation. Also Patrick Bruno is thanked for useful and stimulating discussions.


\end{document}